\documentclass[%
 reprint,
superscriptaddress,
 amsmath,amssymb,
 aip,
]{revtex4-2}

\usepackage{graphicx}% Include figure files
\usepackage{dcolumn}% Align table columns on decimal point
\usepackage{bm}% bold math
\usepackage{orcidlink}
\usepackage{hyperref}
\usepackage{multirow}
\usepackage{mathptmx}
\usepackage{textcomp}

\begin{document}

\newcommand{\beginsupplement}{%
        \setcounter{table}{0}
        \renewcommand{\thetable}{S\arabic{table}}%
        \setcounter{figure}{0}
        \renewcommand{\thefigure}{S\arabic{figure}}%
     }

\preprint{APS/123-QED}

\title{Towards the reproducible fabrication of conductive ferroelectric domain walls into lithium niobate bulk single crystals}% Force line breaks with \\

\author{Julius Ratzenberger\, \orcidlink{0000-0001-6896-2554}}
    \thanks{These two authors contributed equally to this work.}
    \affiliation{Institut für Angewandte Physik, Technische Universität Dresden, 01062 Dresden, Germany}
\author{Iuliia Kiseleva\, \orcidlink{0009-0002-5435-056X}}%
    \thanks{These two authors contributed equally to this work.}
    \affiliation{Institut für Angewandte Physik, Technische Universität Dresden, 01062 Dresden, Germany}
\author{Boris Koppitz\, \orcidlink{0009-0001-6586-0947}}
    \affiliation{Institut für Angewandte Physik, Technische Universität Dresden, 01062 Dresden, Germany}
\author{Elke Beyreuther\,\orcidlink{0000-0003-1899-603X}}
    \email{elke.beyreuther@tu-dresden.de}
    \affiliation{Institut für Angewandte Physik, Technische Universität Dresden, 01062 Dresden, Germany}
\author{Manuel~Zahn\,\orcidlink{0000-0003-0739-3049}}
    \affiliation{Institut für Angewandte Physik, Technische Universität Dresden, 01062 Dresden, Germany}
    \affiliation{Experimental Physics V, Center for Electronic Correlations and Magnetism, University of Augsburg, 86159~Augsburg, Germany}
\author{Joshua~Gössel}
    \affiliation{Institut für Angewandte Physik, Technische Universität Dresden, 01062 Dresden, Germany}
\author{Peter~A.~Hegarty\, \orcidlink{0000-0002-9882-9866}}
    \affiliation{Institut für Angewandte Physik, Technische Universität Dresden, 01062 Dresden, Germany}
\author{Zeeshan H. Amber\, \orcidlink{0000-0002-1796-4979}}
    \affiliation{Institut für Angewandte Physik, Technische Universität Dresden, 01062 Dresden, Germany}
\author{Michael~Rüsing\,\orcidlink{0000-0003-4682-4577}}
     \affiliation{Paderborn University, Institute for Photonic Quantum Systems, 33098 Paderborn, Germany}
\author{Lukas M. Eng\,\orcidlink{0000-0002-2484-4158}}
    \affiliation{Institut für Angewandte Physik, Technische Universität Dresden, 01062 Dresden, Germany}
    \affiliation{ct.qmat: Dresden-Würzburg Cluster of Excellence—EXC 2147, Technische Universit\"at Dresden, 01062 Dresden, Germany}
\date{\today}% It is always \today, today,

\begin{abstract}
Ferroelectric domain walls (DWs) are promising structures for assembling future nano-electronic circuit elements on a larger scale, since reporting domain wall currents of up to 1~mA per single DW. One key requirement hereto is their reproducible manufacturing by gaining preparative control over domain size and domain wall conductivity (DWC). To date, most works on DWC have focused on exploring the fundamental electrical properties of individual DWs within single shot experiments, with emphasis on quantifying the origins for DWC. Very few reports exist when it comes to compare the DWC properties between two separate DWs, and literally nothing exists where issues of reproducibility in DWC devices have been addressed. To fill this gap while facing the challenge of finding guidelines achieving predictable DWC performance, we report on a procedure that allows us to reproducibly prepare single hexagonal domains of a predefined diameter into uniaxial ferroelectric (FE) lithium niobate (LN) single crystals of 200 and 300~\textmu m thickness, respectively. We show that the domain diameter can be controlled with an error of a few percent. As-grown DWs are then subjected to a standard procedure of current-controlled high-voltage DWC enhancement, repetitively reaching a DWC increase of 6 orders of magnitude. While all resulting DWs show significantly enhanced DWC values, subtle features in their individual current-voltage (I-V) characteristics hint towards different 3D shapes into the bulk, with variations probably reflecting local heterogeneities by defects, DW pinning, and surface-near DW inclination, which seem to have a larger impact than expected.
\end{abstract}

%\keywords{Suggested keywords}%Use show keys class option if keyword
\maketitle

\section{Introduction}

In recent decades, lithium niobate (LiNbO$_3$, LN) has become a subject of intense research and application in various fields, ranging from ferroelectric random-access memories \cite{Waser2004} and rectifying junctions \cite{Zhang2021, Qian2022, Suna2023} to memristors \cite{Kaempfe2020, Chaudhary2020}, transistors \cite{McConville2020, Sun2022}, as well as the wide range of photonic and optical devices \cite{Lin2020, Boes2018, Zhang052021}. This popularity stems from LN's high Curie temperature, low optical damage, and commercial availability, since usually grown by the Czochralski method \cite{Schroeder2012}. With such a versatility, stability, and accessibility, LN is often referred to as representing the "silicon of photonics" \cite{Reichenbach2014}.

Apart from its importance in photonics, one of the most captivating aspects of LN lies in its domain wall conductance (DWC), which can be induced purposely at the boundaries between ferroelectric (FE) domains. Domain walls (DWs) display unique electrical properties, with conductivities reaching values that are orders of magnitudes larger as compared to the surrounding bulk material \cite{Godau2017, Werner2017, Zahn2023}. This characteristic makes DWs promising candidates for exploration in nanoelectronics, offering opportunities for rewritable electronics \cite{Stolichnov2015}, applications as components in neuromorphic electronics \cite{Suna2022}, or to assemble faster and energy-efficient electronic components \cite{Xia2019}. Their responsiveness to external stimuli, such as electric fields \cite{Godau2017,KGWB2019} or mechanical strain fields~\cite{singh2022}, opens up the door to engineering DWC-based devices, including non-volatile memories~\cite{Waser2004}, memristors~\cite{Kaempfe2020, Chaudhary2020}, \emph{pn}-junctions\cite{Qian2022,Maguire2023}, and transistors~\cite{McConville2020, Sun2022}.

Nonetheless, despite this huge potential and bright perspective, pinning down all factors that influence the charge transport along DWs and then reproducibly engineer such charged DWs (CDWs) into advanced devices, is hindered on one hand by the lack in understanding these underlying physical mechanisms, while on the other hand methods and protocols of how to then manufacture identical CDW objects, are missing. Luckily, some works have recently (i) addressed the question of \emph{which transport phenomena} are relevant for DWC \cite{Wolba2018, Xiao2018, Zahn2023}, (ii) measured typical activation energies \cite{Zahn2023}, and (iii) extracted charge carrier-types by different methods ranging from magneto-electric resistance \cite{Gregg2022} towards two-terminal AFM-tip based \cite{cam16,tur18} and standard four-point probe Hall measurements \cite{Beccard2022, Beccard2023}.

To investigate CDWs, methods that allow for the local generation of domains are required. Notably, in many experimental applications the preferred tools hereto are AFM-tip writing and UV-assisted poling. While the first technique is preferentially selected for writing domains into thinner films, the latter allows creating singular domains (and walls) into macroscopically thick crystals that measure hundreds of \textmu m in thickness. Such bulk single crystals are ideal model systems to study the underlying physics of DWC, since well discriminating between interface-related and DW-induced transport effects \cite{Zahn2023}. Notably, some literature exists reporting on the method of laser-assisted CDW production in LN, which is able to significantly lower the electric field \cite{Sones2005, Wengler2004, Steigerwald2010} as applied for local poling; nonetheless, \emph{systematic} investigations of the exact correlation between all relevant process parameters and the resulting domain size and structure are missing to date. However, such a lack of reproducibility introduces uncertainties in device characteristics, impacting reliability, functionality, and performance, clearly not matching industry standards. Therefore, addressing the reproducibility challenge is vital to fully harness the potential of CDWs, and to push advancements in FE-based devices.

In this article, we embark on a comprehensive investigation of more than 60 samples having CDWs engineered into 200- and 300-\textmu m-thick single crystals of 5-mol-$\%$ MgO-doped LiNbO$_3$ (LN), in order to develop a poling protocol and procedure that allows to achieve a reproducible fabrication of CDWs. We provide a detailed description of a home-built, automated, computer-controlled setup for domain fabrication via UV-assisted liquid-electrode poling, and analyze factors influencing the domain formation, in particular their area $A_d$. These include the applied electric field $E$ upon poling, the field exposure time $t_p$, and the substrate's chemical composition through the comparison of different LN wafers. Based on these findings, we propose a standardized protocol to facilitate reproducible domain creation. Furthermore, we explore and expand on a process, first developed by Godau \emph{et al.}~\cite{Godau2017}, to \emph{enhance} the conductivity of as-created DWs by many orders of magnitude via high-voltage (HV) treatment, delving deeper into that effect through statistical analysis of a large number of identically-prepared samples. We study the impact of the above-described factors during the poling process as well as the influence of the HV parameters during DWC-enhancement, providing valuable insights and suggestions for further process improvements.

\section{Materials and Methods}

\subsection{Samples}

LN is commercially available as an uniaxial FE material with a large optical bandgap of $E_{gap}=4.0$~ eV ~\cite{Schroeder2012}. Its singular polarization axis runs parallel to the crystallographic z- or c-axis. In the following, we will refer to the two different polarization directions and the respective terminating surfaces as z- and z+. All samples used in this study measure typically 5$\times$6~mm$^2$ in x-y-direction, and are cut-out from three different wafers supplied by \emph{Yamaju Ceramics Co., Ltd. (Japan)}, all with the same nominal doping concentration of 5-mol\%~MgO. In detail, these include two z-cut LN wafers of a 200~\textmu m thickness each, and purchased one year apart from each other (2019 and 2020). Furthermore, in 2022 we acquired a third z-cut LN wafer from the same company, being 300~\textmu m thick. Naturally, these three wafers were produced from three distinct boules in different years, and hence allow us to study whether or not domain growth and DWC might change when comparing wafers from one and the same manufacturer. Note, that we and others \cite{Wengler2004} had observed that wafers from different sources may show slight variations in domain growth, e.g., in the domain propagation velocity, despite of nominally having the same specifications, i.e., 5-mol-$\%$ MgO-doping.

Throughout this publication, we label all samples that contain DWs as A-B-C, with A = 1,2,3 declaring the three wafers purchased in 2019, 2020, and 2022, respectively, B = 200 or 300 being the wafer thickness (in ~\textmu m), and C being the consecutive sample serial number during our investigations. As an example, "sample~1-200-3" refers to the third sample from wafer~1, with a thickness of 200~\textmu m. Furthermore, the samples are grouped within so-called \emph{batches}, containing 6-8 specimens per batch; each batch allows for comparison with respect to the concrete influence of one or more process parameters on the domain area and/or the final current-voltage (I-V) characteristics. All samples host one single hexagonal FE domain and its respective DWs, the latter being prepared under varying conditions as specified in Tab.~\ref{tab:samples} (quick guide) and Sec.~A of the supplementary material (full-length table containing all 63 single samples of this work) and explained in the following paragraphs in detail. 

\begin{table*}
\caption{\label{tab:samples}Short overview of the sample batches of this study, the respective domain fabrication ("poling") parameters, and -- if applicable -- parameters of the domain wall conductivity (DWC) enhancement process. The table contains also information on whether I-V curves after domain growth and after "enhancement" were captured or not (symbolized by "$+$" or "$-$", respectively), as well as the specific type of investigation(s), for which every sample batch had been assigned. A more detailed list containing every individual sample can be found in the supplementary material (Sec.~A). Note that batches 1.4 and 1.5, which were used only within preliminary experiments exclusively dedicated to finding the most reasonable laser power and NaCl concentration for the liquid electrodes, are omitted here.}
\begin{ruledtabular}
\renewcommand{\arraystretch}{1.3}
\begin{tabular}{llcccc}
{\bf Batch} & {\bf Labels} & {\bf Poling} & {\bf DWC-enhancement} &  {\bf I-V-data} & {\bf Purpose(s)}\\
    &   & {\bf parameters} & {\bf parameters} &     & {\bf of investigation}\\
\hline\hline

1.1 & 1-200-1$\ldots$8 & $E=4.0$~kV/mm; & -- & $--$ 
 & reproducibility of $A_d$\\
 &  & $t_p=$30~s &  & 
 & \\ \hline
1.2 & 1-200-9$\ldots$15 & $E=4.0$~kV/mm; & $v=$4~V/s;  & $++$ & $A_d=f(t_p)$~@~$E=$4.0~kV/mm; \\
 &  & $t_p=$10-180~s \footnotemark[1] & $V_{max}$=500~V &  &  reproducibility of final I-V curves \\ \hline 
 1.3 & 1-200-16$\ldots$22 & $E=4.5$~kV/mm; & $v=$4~V/s;  & $++$ & $A_d=f(t_p)$~@~$E=$4.5~kV/mm; \\
 &  & $t_p=$10--180~s \footnotemark[1] & $V_{max}$=500~V &  &  reproducibility of final I-V curves\\ \hline\hline 
2.1 & 2-200-1$\ldots$6 & $E=4.0$~kV/mm; & $v=$4~V/s;  & $++$ & reproducibility of $A_d$; \\
 &  & $t_p=$120~s & $I_{max}$=1~\textmu A &  &  reproducibility of final I-V curves \\ \hline 
2.2 & 2-200-7$\ldots$13 & $E=4.5$~kV/mm; & $v=$4~V/s;  & $++$ & $A_d=f(t_p)$~@~$E=$4.5~kV/mm; \\
 &  & $t_p=$10-180~s \footnotemark[1] & $V_{max}$=500~V &  &  reproducibility of final I-V curves \\
 \hline\hline 
3.1 & 3-300-1$\ldots$6 & $E=4.67$~kV/mm; & $v=$4~V/s;  & $++$ & reproducibility of final I-V curves \\
 &  & $t_p=$120~s & $I_{max}$=1~\textmu A &  &   \\ \hline 
3.2 & 3-300-7$\ldots$13 & $E=4.67$~kV/mm; & --  & $+-$ & $A_d=f(t_p)$~@~$E=$4.67~kV/mm \\
 &  & $t_p=$10-180~s \footnotemark[1] &  &  &   \\
\end{tabular}
\end{ruledtabular}
\renewcommand{\arraystretch}{1.0}
\footnotetext[1]{Concrete $t_p$-values: 10; 20; 30; 60; 90; 120; 180~s.}
\end{table*}

\subsection{Growing Hexagonally-Shaped Domains by UV-Assisted Liquid-Electrode Poling}

\begin{figure*}[!htb]
\includegraphics[width=\textwidth]{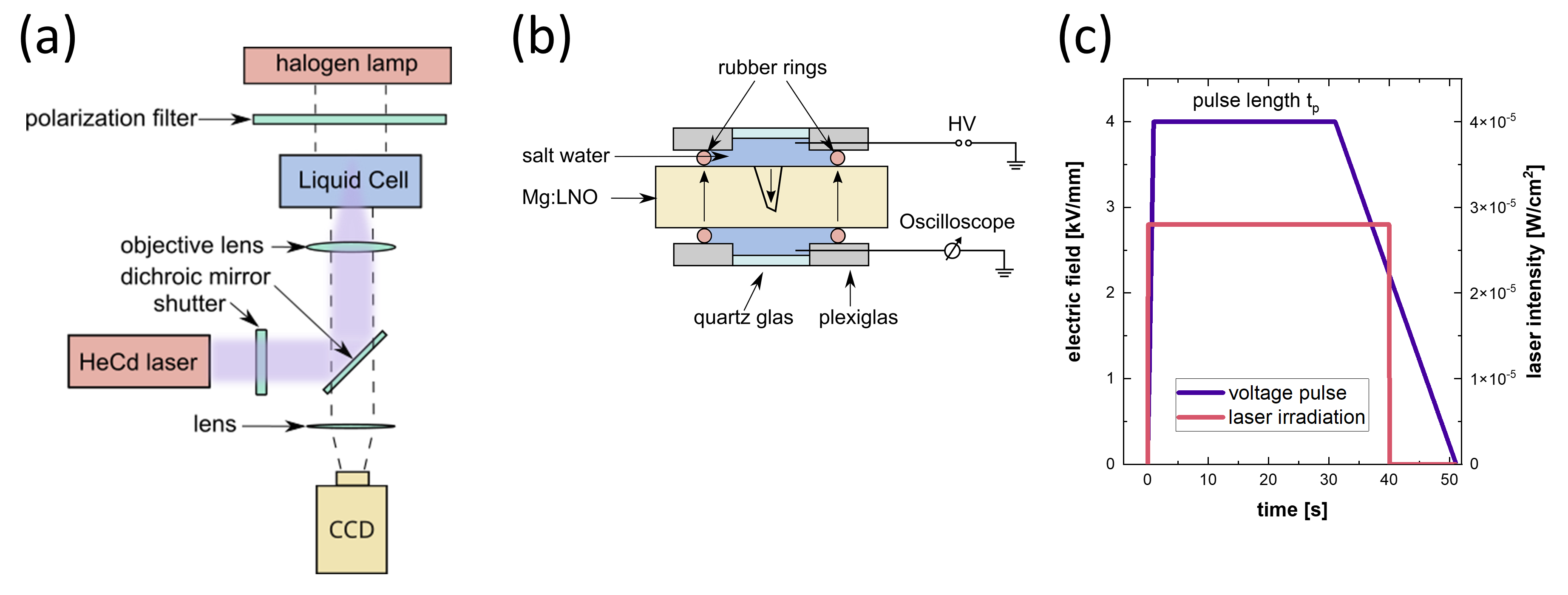}
\caption{\label{fig:setup} Schematics illustrating the growth of hexagonal ferroelectric domains in LiNbO$_3$ single crystals: (a) Overview picture of the optical setup: A HeCd laser with a wavelength of 325~nm acts as the UV-light source. Utilizing various optical lenses and an inverted microscope, the UV light is focused on the top of the sample in a home-built liquid cell. The laser exposure time is precisely controlled by a computer-controlled shutter. A halogen lamp with a linear polarization filter illuminates a CCD camera through the crystal, allowing in-situ observation of the growing domains. (b) Sketch of the liquid cell with the electronic circuit for monitoring and adjusting an applied electric field. (c) Diagram of an exemplary poling recipe, showing the poling field (violet line) and laser irradiation (red line) applied to the samples. Immediately upon the start of a poling procedure, the laser irradiation begins at $2.8\times10^{-5}$~W/cm$^{2}$, and the electric field $E$ is increased to 4~kV/mm over a period of 1~second. For the next period -- $t_{p}$ =30~s -- the electric field and the laser irradiation are kept constant. Afterward, the electric field is slowly, linearly decreased to zero over $t_{rampdown}$=20~s to minimize any spontaneous back-switching effects. During this process, the laser irradiation is switched off when the electric field reaches half of its original value.}
\end{figure*}

We constructed a UV-light-assisted poling setup to grow our domain structures, as depicted in Fig.~\ref{fig:setup}(a). During a typical domain poling procedure, the 325-nm HeCd-laser (\emph{Kimmon Koha IK3301R-G}) is focused onto the z+~crystal surface using an apochromatic lens with NA = 0.3 to a spot size of around 4.5~\textmu m in diameter. At this wavelength, the crystals exhibit a low absorption coefficient \cite{Wengler2005} of $\alpha$ = 0.5~mm$^{-1}$, resulting in a $1/e$-penetration depth of 2~mm, which hence implies full penetration across all samples. The incident power was kept constant at $2.8\times10^{-5}$~W/cm$^{2}$ for all experiments; that value was chosen as the favorable power value after having run a separate investigation beforehand to evaluate the impact of the applied laser beam power on the resulting domain area $A_d$ by using sample batch~1.4 (see Sec.~G of the supplementary material for details).

The 325-nm-UV-laser light generates an influx of charge carriers at the illuminated spot and thus lowers the coercive field locally significantly to only 20\% of the value in darkness (the general phenomenon had been reported earlier~\cite{Steigerwald2010}). Therefore, a significantly lower electric field of 3.5~kV/mm at the spot is required to initiate the inversion of the polarization. This field was generated by an \emph{Agilent 33220A} arbitrary waveform generator, whose output was amplified by two different HV amplifiers: (a) A \emph{Matsosada Precision AMT-20B10-LC(230 V)} HV amplifier with a maximum output current of $\pm$10~mA, a maximum voltage of 20~kV, and a slew rate of up to 360~V/\textmu s, and (b) using a \emph{Trek 2210} with a maximum output current of $\pm$20~mA, a maximum voltage of 1~kV, and a slew rate of up to 150~V/\textmu s. We noticed no differences in the poling process and kinetics in the overlapping voltage range of the two HV amplifiers. Additionally, we monitored the voltage pulse as well as the poling current by a \emph{Tektronix TDS2024B} digital oscilloscope. 

\begin{figure}%[!htb]
\includegraphics[width=0.45\textwidth]{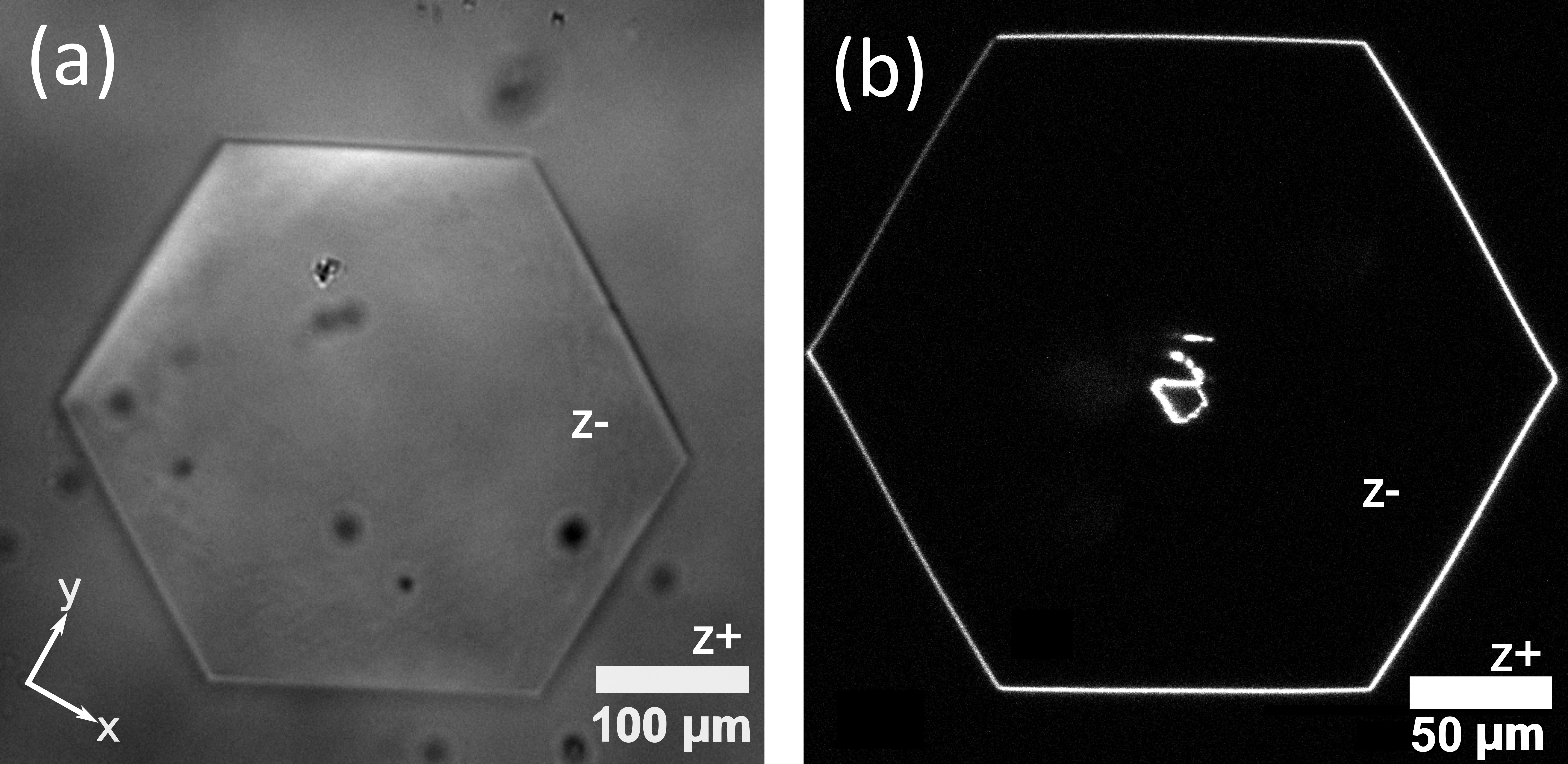}
\caption{\label{fig:SHGvsPLM} Image of the domain in sample 2-200-17 under (a) a polarization-sensitive light microscope (PLM), and (b) a second harmonic generation microscope (SGHM). For further details on both imaging methods and the extraction of the domain area, see Secs.~B to E in the supplementary material.}
\end{figure}

In addition, the \emph{in-situ} nucleation and growth of domains was monitored in real-time using a white-light polarization microscope illuminating the sample from the top and a CCD camera. Diffraction-limited domain tracking thus was possible due to the strain-induced birefringence close to DWs. Samples were fixed into the microscope in a custom-built liquid cell sample holder, sketched in Fig.~\ref{fig:setup}(b). The liquid cell itself consists of two plexiglas parts having a quartz glass window in their center, to ensure the optical transmission of UV light. The sample was held in place between two rubber O-rings acting as spacers, while the surrounding cell volume on both sample sides was filled with a 2-wt$\%$ NaCl solution. These "liquid electrodes" ensure transparency while providing a uniform electric field distribution across the whole sample surface~\cite{Chen2003}. An analysis of domain area $A_d$ as a function of NaCl concentration was performed in a second preliminary experiment that is reported in the Sec.~H of eth supplementary material, using sample batch~1.5, and showing the independence on $A_d$ over a broad range of NaCl-concentrations.   

To ensure reproducibility, the poling experiments were always conducted by following the same sequences of sample treatment:
\begin{itemize}
    \item {\bf (i)} A fresh sample was extracted from a LN wafer, and 
    \item {\bf (ii)} subjected to a thorough cleaning procedure that included a 10-minutes plasma etching step followed by a 5-minutes ultrasonic cleaning in each acetone and isopropanol, and finally rinsed with deionized water.
    \item {\bf (iii)} Each sample was then mounted into the liquid NaCl-cell for domain poling, with the z+ side always facing upwards. Equally, the laser always entered the sample by the z- side and was focused onto the z+ interface. 
    \item {\bf (iv)} The subsequent \emph{poling protocol} is shown in Fig.~\ref{fig:setup}(c): At the beginning, a sharp voltage ramp was applied lasting for one second until the maximum electric field is reached. Then the selected maximum poling voltage was kept constant for a selected duration, i.e. for a certain pulse length $t_{p}$. To terminate the poling process, the electric field is slowly lowered over a time period of $t_{rampdown}$ = 20~s to minimize the effects of spontaneous back-switching \cite{Kim2001}. Throughout the poling process, the sample was constantly irradiated till 10~seconds to the end, as the applied electric field reached a value lower than half of the coercive field $E_c$. 
    \item {\bf (v)} Every freshly grown domain is subsequently imaged by polarization light microscopy (PLM) and the grown domain area be extracted from measurements of the hexagon's side lengths. In selected cases (samples 01-03 of batch 2.1 and all samples of batches 1.2, 1.3, 2.2, 3.1, and 3.2.), 3D images of the full DW morphology were collected by second-harmonic generation microscopy (SHGM). An elaborate comparative description of both imaging methods (see Fig.~\ref{fig:SHGvsPLM} for typical results) as well as the mathematical details on domain area determination are given in Secs.~B--E of the supplementary material. 
    \item {\bf (vi)} After having acquired the PLM images of as-freshly-formed domains, all samples underwent the above cleaning protocol once more, however, starting with an initial deionized-water rinsing step to eliminate any salt residues. 
    \item {\bf (vii)} Finally, 10-nm-thin chromium electrodes were vapor-deposited under high-vacuum conditions (base pressure of 10$^{-6}$~mbar) onto both sides of the LN crystals using a 3~mm~×~3~mm shadow mask that fully covers the newly-grown domain. This insures proper contacts to DWs for both current-voltage (I-V) recording and DWC enhancement through high-voltage (HV) ramping. 
\end{itemize}

\subsection{Current-Voltage (I-V) Characterization and DWC Enhancement Procedure}
\label{sec:methods:CDW_enh}

The vision for the investigated DWs is their application in various electronic commercial devices -- thus the current-voltage (I-V) dependence and its predictability are the clue characteristics to be evaluated. In this work, we recorded I-V curves of all DWs at all stages of production. The above-mentioned Cr electrodes were therefore connected to a \emph{Keithley 6517B} electrometer using metal wires and conductive silver paint. The electrometer was used twofold, first to recording the macroscopic I-V data, and second as the HV source to carry out the DWC enhancement procedure. I-V curves between -10~V to $+$10~V and voltage increments of $\delta$V = 0.5~V with $\delta$t = 2~s were acquired both before and after enhancement.

DWC enhancement was conducted based on the protocol as introduced by Godau \emph{et al.}~\cite{Godau2017}: the –z-electrode of every sample is set to ground, and a negative voltage is applied  to the $+$z-side increasing linearly at a rate of $v$ = 4~V/s up to a value $V_{max}$ that corresponds to 60\% of $E_{C}$, hence up to ~3.3 kV/mm~\cite{Kirbus2019}. In the following, the as-summarized DWC-enhancement procedure will be referred to as \emph{voltage control}. Initially, i.e., for batches 1.2, 1.3, and 2.2, voltage ramps up to $V_{max}$ = 500~V were applied, but this led to the disintegration of the original domain structure into smaller fragmented regions containing numerous tiny needle-like spike domains, named hereafter as "exploded" domains; we had vastly investigated those type of domains being described for a single LN sample by Kirbus \emph{et al.}~\cite{Kirbus2019} earlier. Notably, during this procedure, a sudden current jump from $10^{-5}$A to $10^{-3}$A was observed.

To address this issue, we managed to improve the DWC-enhancement procedure, by stopping the voltage ramping whenever the maximum current $I_{max}$ of 10$^{-6}$A is reached, and then keep this corresponding voltage value; accordingly, this type of DWC-enhancement is labelled \emph{current control} hereafter. Notably, \emph{current control} preserves the original domain structure and size in the majority of cases investigated here. All subsequent DWC enhancement procedures were hence carried out using this improved approach, in particular with sample batches 2.1 and 3.1. 

As a quick overview, Tab.~\ref{tab:samples} displays a listing for which sample batch I-V-data have been acquired, as well as whether or not, and under which circumstances (stopping by reaching $V_{max}$ or $I_{max}$) the "enhancement" procedure was conducted.

\section{Results and discussion}
\subsection{Reproducibility and Tunability of the As-Grown Domain Area}

In this part, the domain engineering method of UV-assisted liquid-electrode poling is investigated with respect to the impact of several process parameters (poling pulse duration $t_p$, electrical field strength $E$, wafer type) on the final domain area $A_d$ and the I-V-characteristics of its DW, while a detailed description of the growth process itself, as (i) known from the literature and (ii) derived from a SHGM investigation of the DW inclination as a function of the poling pulse length $t_p$, is summarized in Sec.~F of the supplementary material.

\begin{figure*}[!htb]
\includegraphics[width=\textwidth]{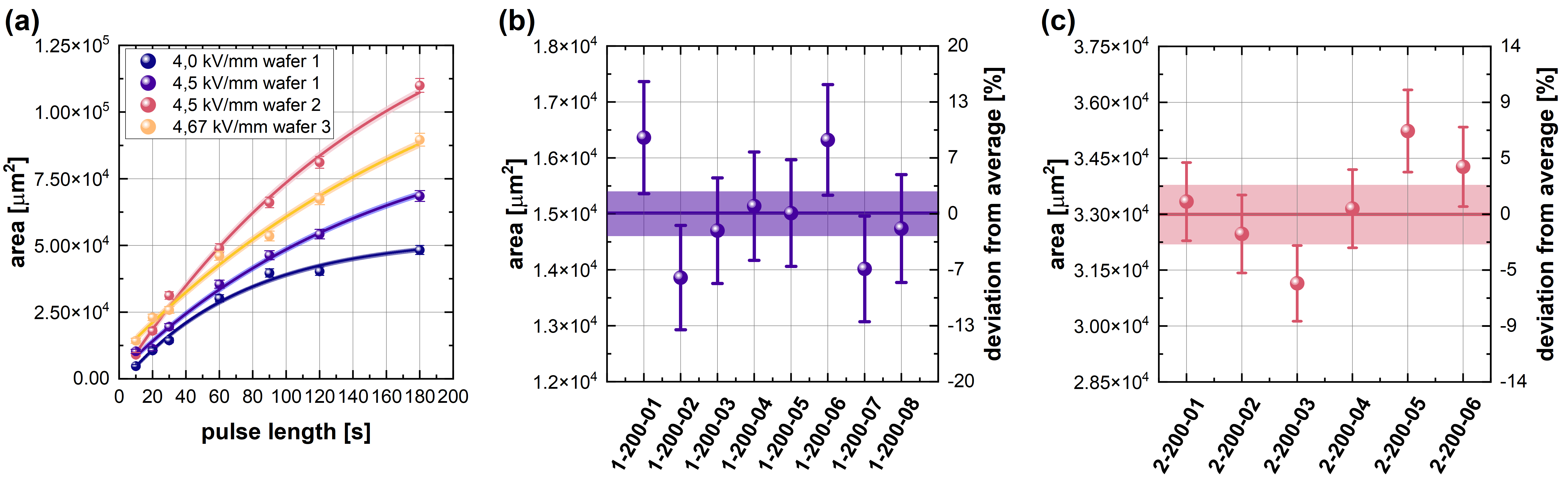}
\caption{\label{fig:comparison} 
(a) To investigate the dependence of the domain area on the poling-voltage pulse duration, we varied $t_{p}$ between 10~s and 180~s. This resulted in an increase of the domain area of one order of magnitude between the shortest and longest pulse. Furthermore, three different electric fields, as well as samples from three different wafers, were investigated. A noticeable size difference was found between the sample batches of the different wafers. (b,c)
To probe the reproducibility of the created domain area, two measurement series of domains fabricated with the same poling recipe were carried out. One series with $t_{p}$=30~s for samples from wafer~1 in (b) and another series with $t_{p}$=30~s for samples originating from wafer~2 in (c). The orange or blue lines indicate the mean area reached with this particular poling protocol, respectively.}
\end{figure*}

\subsubsection{Domain Area as a Function of Poling Pulse Duration, Electrical Field, and Wafer Type}

To gain an improved understanding of the domain areas' dependence on the poling time $t_p$, we performed several series of measurements with $t_p$ being varied between 10 and 180~seconds for the sample batches 1.2 (dark blue), 1.3 (violet), 2.2 (red), and 3.2 (gold), as plotted in Fig.~\ref{fig:comparison}(a). In addition, we varied the electric field strength $E$ and compared samples between the three LN wafers. The key findings of these comparative measurements are described in the following.

Regardless of the field strength or the wafer, the dependence of the poling duration $t_p$ on the domain area always shows the same trend: a sharp increase in size at short poling times slowly flattens out the longer the poling process takes. This behavior can be attributed to the laser irradiation -- the area surrounding the laser spot maintains a significantly lower coercive field as compared to distant regions, facilitating a faster growth rate and DW propagation speed. When the coercive field increases due to a lower number of free charge carriers in less illuminated areas, the expansion slows down until it reaches a stable growth rate. As reported in the literature, the growth rates in the undisturbed regions of the crystal were steady at a constant electric field \cite{Gopalan1999,Shur2006}.

To further investigate growth parameters, the same measurement series was carried out with an increased electric field of 4.5~kV/mm. This led to a faster domain growth and the domain area increased 42$\%$ in comparison to the previous measurement series on the same wafer. The overall trend did not change, with a slowing down of the expansion speed as the poling length increased.

The same measurement series was then performed additionally with samples obtained from wafer~2 and wafer~3 with similar electric field strength, as also pictured in Fig.~\ref{fig:comparison}(a). Surprisingly, we found significant differences in growth rates between all three wafers. The samples from wafer~2, red dots in Fig~\ref{fig:comparison}(a), exhibited the fastest growth. In contrast, the samples from wafer~1 (violet) grew slowest, even though they underwent the same poling protocol. The thicker samples from wafer~3 (gold) were poled with a slightly higher electric field but still reached a similar growth speed as the other specimens. We attribute these differences to small changes in defects and impurities leading to different amounts of pinning events for DWs and thus different growth rates~\cite{Kim2001}. This also means that a careful \emph{calibration} of growth speeds is required to achieve domains of comparable sizes, in particular when samples are produced from different wafers -- even if the latter have nominally the same chemical composition.

\subsubsection{Size Variations for Virtually Equally Grown Hexagonal Domains}

The goal of these experiments was to check the reproducibility of the domain size within the UV-assisted liquid-electrode poling process. A series of eight samples from wafer 1 with a $t_{p}$~=~30~s (batch~1.1) and six samples from wafer 2 with a $t_{p}$~=~120~s (batch~2.1) were prepared and the domain areas exemplarily compared, as shown in Fig.~\ref{fig:comparison}(b,c). The shorter pulse duration $t_{p}$~=~30~s, as illustrated in Fig.~\ref{fig:comparison}(b), resulted in a smaller area of 1.5$\times10^{4}$~\textmu m$^{2}$ with a deviation of 7$\%$. In contrast, the longer pulse duration, as shown in Fig.~\ref{fig:comparison}(c), yielded an area of 3.25$\times10^{4}$~\textmu m$^{2}$ with a deviation of 9$\%$. All domains exhibited a hexagonal shape, as depicted in Fig.~\ref{fig:SHGvsPLM}, characterized by one pair of parallel sides that were slightly longer, which can be attributed to the pinning of the domain wall and the slightly oval shape of the laser spot. Overall we conclude that this poling method is capable of creating domains of the same size within a reasonable error. Potentially, variations may be reduced further, if some form of in-situ control of the domain size, i.e., by machine vision via PLM, is applied.

\subsubsection{Electrical Characterization of As-Grown Domain Walls}

\begin{figure*}[!htb]
\includegraphics[width=\textwidth]{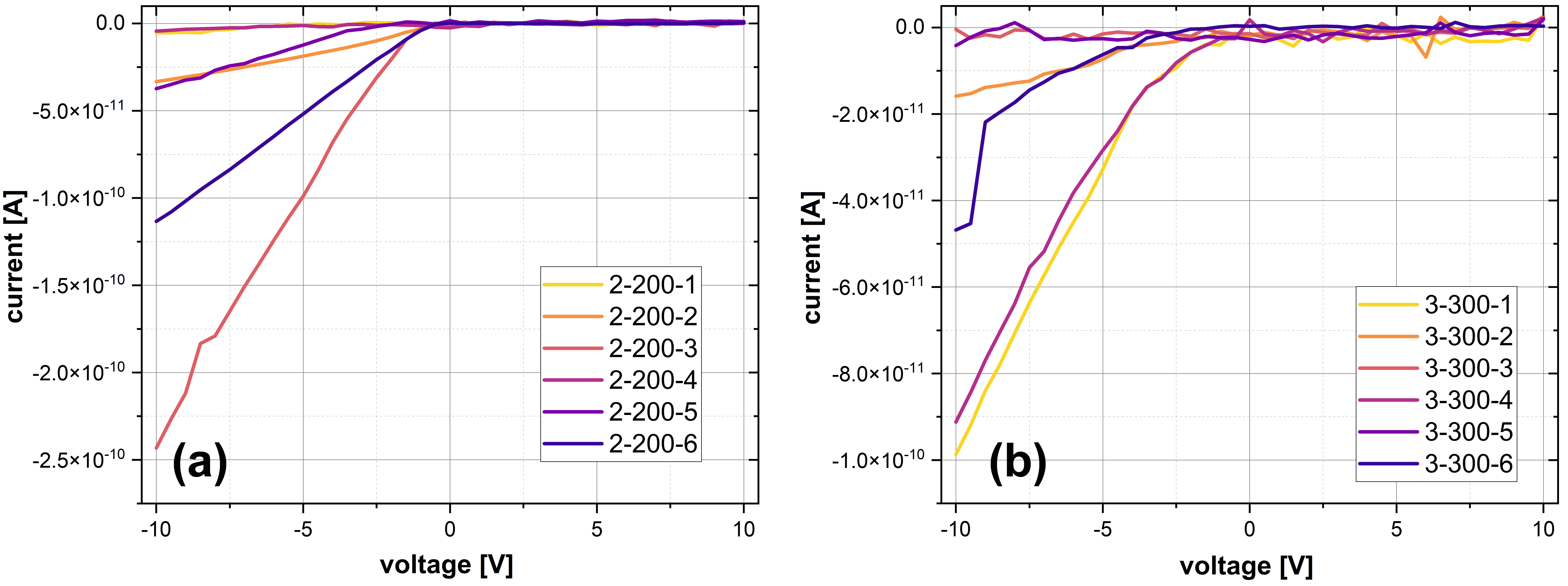}
\caption{\label{fig:initial_conductance} 
Current-voltage curves of the samples from batches 2.1 (a) and 3.1 (b) directly after poling, before any modification of the domain wall structure took place. As can be seen from the pictures, the initial conductance is different for the samples after the poling under identical conditions, which stresses the influence of individual lattice defects in the vicinity of the domain walls and different real structures of the DW-electrode contacts. Currents at negative voltages are significantly higher than at positive ones -- a phenomenon, which has been described previously~\cite{Zahn2023}.}
\end{figure*}

Current-voltage (I-V) characteristics over the $\pm$10-V-range were systematically collected at various stages of conductive DW generation. The first stage is the characterization of the initial as-grown (as-poled) domains. By way of illustration, we present the I-V characteristics of samples obtained immediately after the poling process from batches~2.1 and 3.1 in Fig.~\ref{fig:initial_conductance}. It is worth noting that the two batches exhibit qualitatively similar characteristics. The current-voltage curves display a clear non-linear behavior, asymmetry with respect to voltage polarity (as discussed and explained in ref.~\onlinecite{Zahn2023}), and occasionally pronounced electrical noise due to the low current magnitude; that can be seen especially well in Fig.~\ref{fig:initial_conductance}(b). The maximum absolute current, acquired at $\pm 10$~V for all samples in batches~2.1 and 3.1, falls within the range of 4.3~$\times$10$^{-13}$~A to 2.4~$\times$10$^{-10}$~A for a -10~V bias voltage, and 1.1~$\times10^{-13}$~A to 2.6~$\times10^{-12}$~A at a $+$10~V bias. These values are already very close to the bulk conductivity of LN and correspond to a typical insulator rather than to a conductive domain wall.

Notably, when subjected to the maximum positive voltage, the sample conductance is lower as compared to that under a negative voltage of equivalent absolute magnitude, with a narrower dispersion of values among the samples, as is evident from Figs.~ \ref{fig:initial_conductance}~and~\ref{fig:Current_Change}.

The I-V curves of identically prepared samples prior to DWC "enhancement" exhibit significant variations in magnitude. Given that the production parameters were consistent among all samples, we hypothesize that these electrical disparities may be attributed to local defects and impurities within the crystal lattice, particularly those present near or at the sample surface. This hypothesis is supported by previous findings~\cite{Godau2017}, which established that the geometrical and electronic real structure of the interface between the crystal and electrode serves as a crucially determining factor for charge transport.

\subsection{Reproducibility of the I-V Characteristics After Domain Conductivity Enhancement Procedures}

In this part, the experimental processes of current-voltage characterization and conductivity enhancement of domain walls are examined. This encompasses the application of the protocol proposed by Godau \emph{et al.}~\cite{Godau2017}, the subsequent refinement towards a \emph{current-controlled} approach, and the characterization of the obtained highly conductive DWs (compared to as-poled DWs). 
%The results underscore the complex relationship between material properties and the observed behavior of domain walls, emphasizing the need for further exploration.

\subsubsection{Voltage- vs. Current-Controlled Domain Wall Conductivity Enhancement}

\begin{figure*}[!htb]
\includegraphics[width=0.85\textwidth]{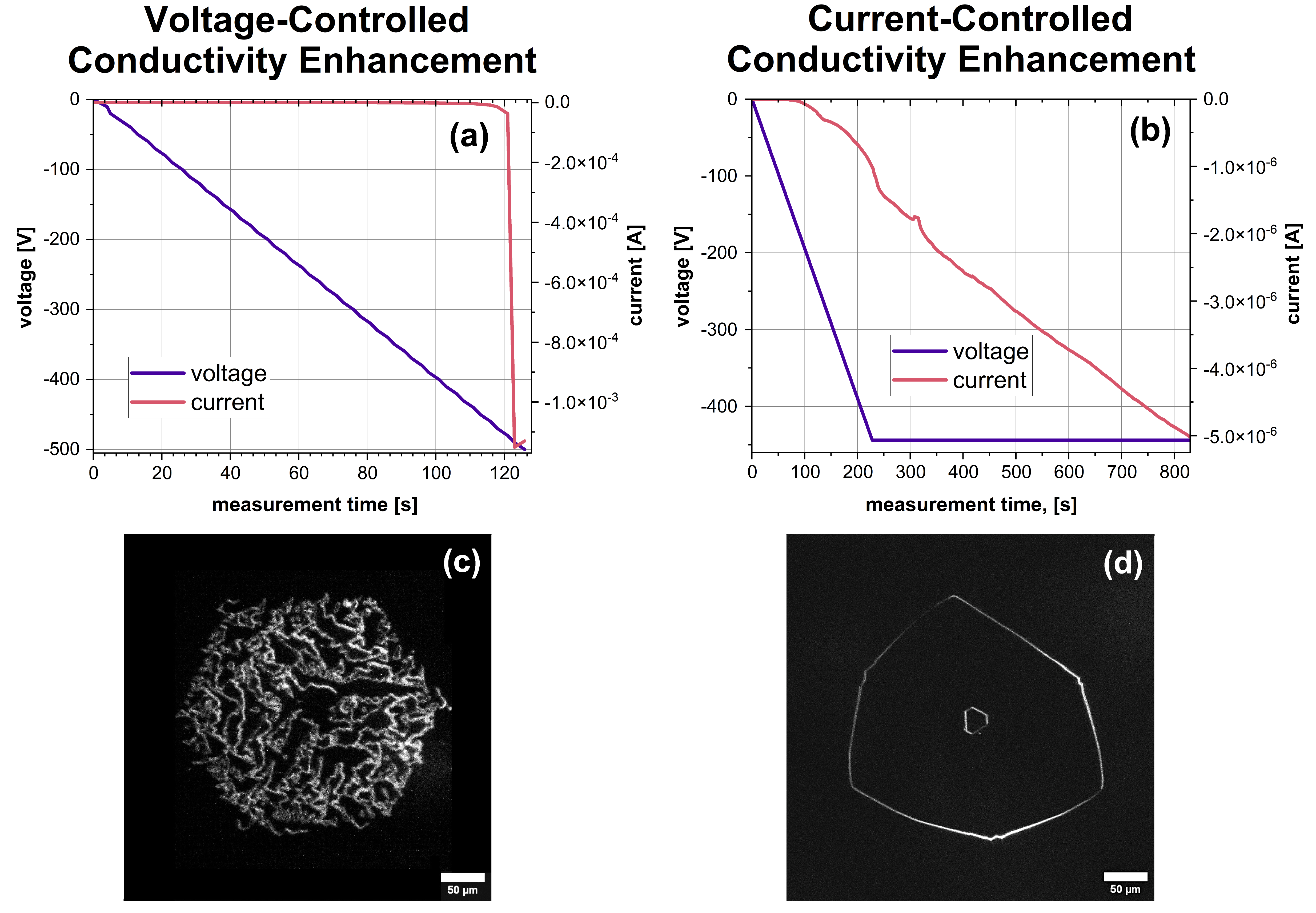}
\caption{\label{fig:typesOfCDWsEnh} 
The I-V characteristics for the two variations of DWC enhancement procedures, i.e., the \emph{voltage-controlled} (a) and the \emph{current-controlled} (b) approach, are illustrated alongside the resulting DW structures [(c) and (d)] of samples 2-200-11 and 3-300-3, respectively. In the first case, a negative voltage is applied to the z+ side of the LN DW sample, gradually increasing at a rate of 4~V/s until reaching 500~V. A subsequent 10-minute stabilization occurs at the maximal voltage. The corresponding plot (a) depicts a swift surge in the current (by more than one order of magnitude) during the voltage ramp-up, resulting in an "imploded" geometrical domain structure (c). This phenomenon was observed across 21 domains of varying sizes with the current rapidly increasing after reaching a value of 10$^{-5}$A in all cases. In an improved version of the procedure (b), the voltage ramp-up at 4~V/s was terminated once the current achieved 10$^{-6}$A, with a 10-minute stabilization period. As evident from plot (b), this modification eliminates the rapid current increase, leading to DWs with a regular geometrical shape (d) in the majority of the cases.}
\end{figure*}

Immediately after I-V characterization of as-poled DWs, the DWC enhancement procedure~\cite{Godau2017}, as described in detail in Sec.~\ref{sec:methods:CDW_enh}, was applied to the samples of the batches~1.2, 1.3, and 2.2. In this original, so-called \emph{voltage-controlled} protocol, the maximal predefined value of the voltage was +500~V  for 200~\textmu m thick samples, followed by a 10-minute stabilization period under the application of the maximal voltage. Throughout this process, a sharp current surge from 10$^{-5}$A to 10$^{-3}$A at 200~V was consistently observed for all samples [exemplified by Fig.~\ref{fig:typesOfCDWsEnh}(a)]. The analysis of recorded SHGM images of these domains unveiled the disintegration of singular hexagonal domains into numerous minute needle-like domains, as depicted in Fig.~\ref{fig:typesOfCDWsEnh}(c). This abrupt current increase, as confirmed by analysis, invariably indicated a domain "implosion", i.e., the fracturing into many needle-like structures, which makes this process uncontrollable and irreversible.

Hence, to achieve a more homogeneous geometrical "enhancement" result, employing the current-controlled scheme was tested for batches 2.1 and 3.1, where each domain wall is voltage-treated until maximum current value is reached only, instead of a predefined target voltage. To achieve this, the increase in voltage was immediately halted as soon as the current reached the level of 10$^{-6}$A, which is an order of magnitude smaller than the critical threshold of 10$^{-5}$A, where the implosion was observed, providing a safety margin. The changes in both current and voltage during the procedure are visible in Fig.~\ref{fig:typesOfCDWsEnh}(b), while the corresponding SHGM image, showcasing an exemplary conductivity-enhanced domain wall, is displayed in Fig.~\ref{fig:typesOfCDWsEnh}(d). It is evident that the current exhibited a rather smooth transition throughout the procedure, resulting in the formation of domain shapes with a regular triangular pattern, as previously described~\cite{Godau2017}. 

\subsubsection{Current-Controlled DWC-Enhancement: Final Electrical Performances and Domain Wall Shapes}

\begin{figure*}[!htb]
\includegraphics[width=0.85\textwidth]{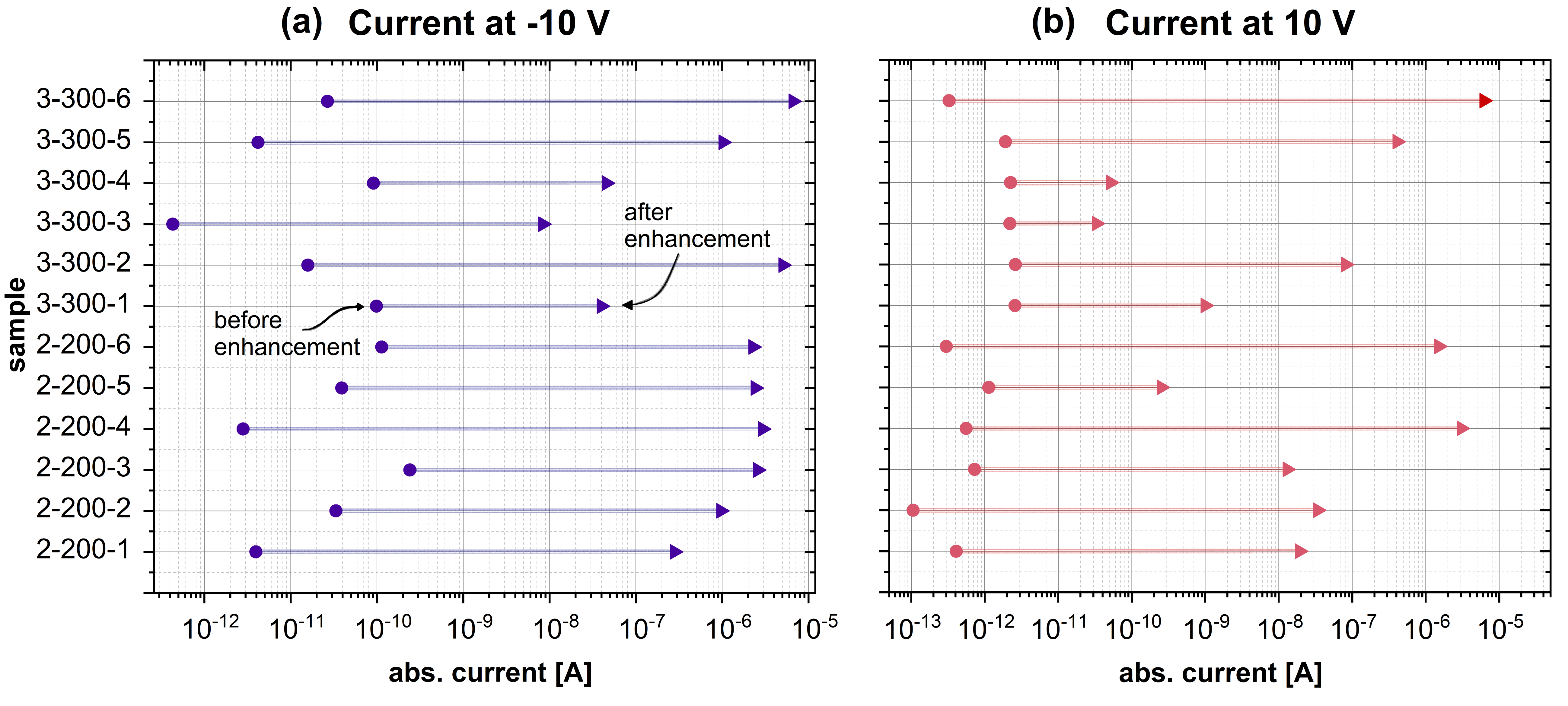}
\caption{\label{fig:Current_Change} Change of the DW current at $\pm 10$~V (red and violet signs, respectively) for the twelve samples of batches 2.1 and 3.1 before (dots) and after (triangles) the DWC enhancement procedure. The magnitudes of current change are spanning from approximately one order of magnitude (sample 3-300-3 at -10~V) to more than seven orders of magnitude (sample 3-300-6 at -10~V). On average, the current at +10~V increases by 5~orders of magnitude and at -10~V by 6~orders of magnitude. For the absolute majority of cases, the current under positive measuring voltage is larger both before- and post-enhancement. The full I-V curves can be found in Fig.~S7 of the supplementary material.}
\end{figure*}

 The -- in terms of geometrical reproducibility favorable -- current-controlled domain wall conductivity enhancement procedure was performed on a statistically significant number of samples: twelve in total. These samples encompass after-poling hexagonal domains of approximately similar sizes of 270~\textmu m in diameter, all of which were poled within the crystals of both wafer~2 (batch~2.1) and wafer~3 (batch~3.1). Although the DWC enhancement procedures were executed under uniform conditions, there exists still a notable distinction in the I-V characteristics observed during and after the enhancement process across all samples. For a visual representation of this variance, refer to Fig.~\ref{fig:Current_Change}, where the DWs are compared based on the maximal currents, obtained at $\pm 10$~V, as well as to Fig.~S7 in the supplementary material, where the corresponding I-V raw data is provided.

\begin{figure*}[!htb]
\includegraphics[width=0.85\textwidth]{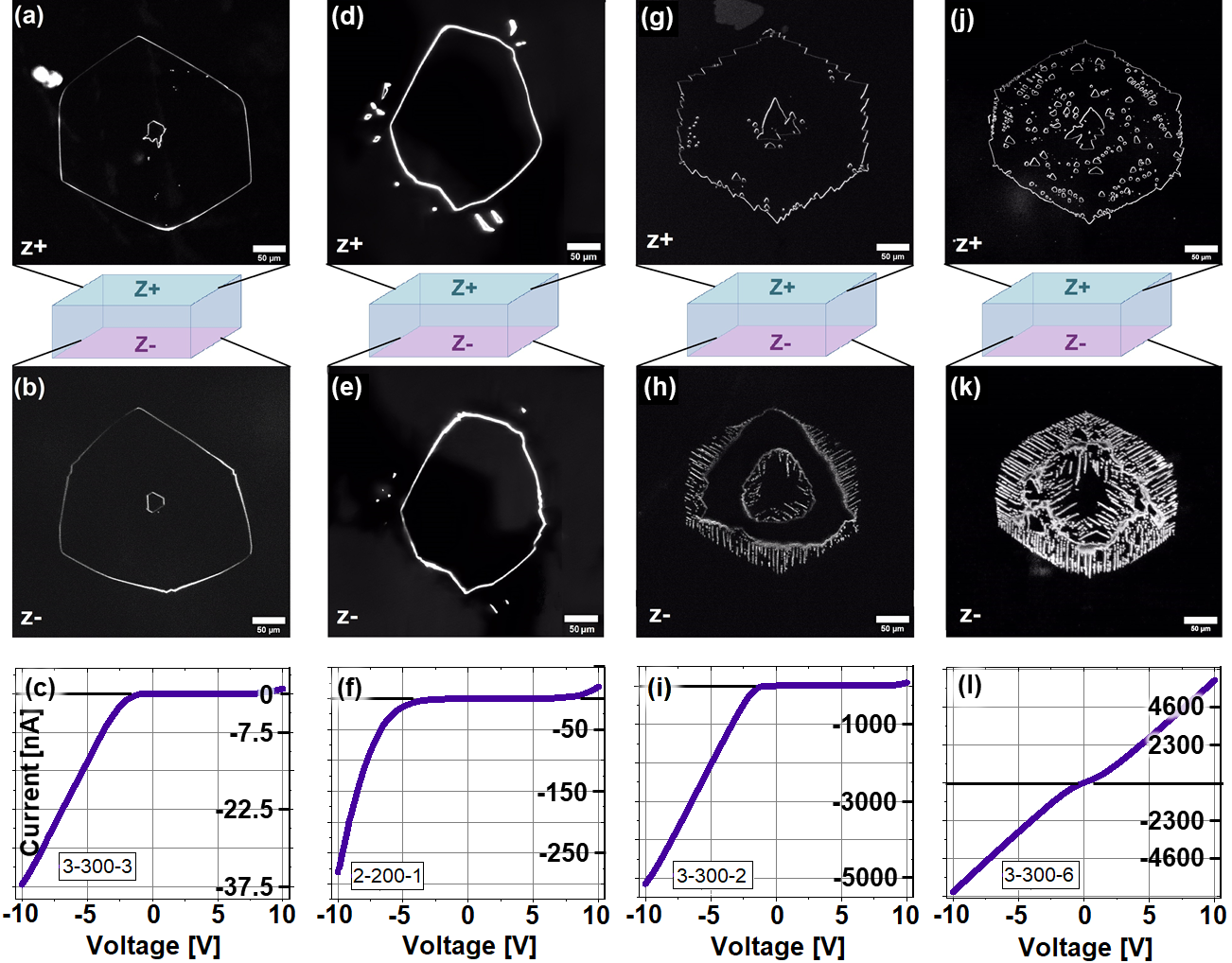}
\caption{\label{fig:SHG_post-enhancement_selection} Domain SGHM images at both +z- and -z- sides of the samples 3-300-3 (a, b), 2-200-1 (d, e), 3-300-2 (g, h), and 3-300-6 (j, k), correspondingly, taken after the DWC-enhancement process using the current-control scheme. As can be seen from the selection of images, the geometrical form of the domains can still vary substantially; from the already described~\cite{Godau2017} 'triangular' shape of the domain like in the case of the sample 1-300-3, to the shape of the irregular oval in case of the sample 2-200-1, where no symmetry in relation to the crystal axes can be traced, towards the cases of the two other samples (3-300-2 and 3-300-6) exhibiting several stages of spike domain formation. In the latter two cases, the residual pinning sites, which appeared during the conductivity enhancement procedure, are clearly visible on the -z-side of the crystal (and through all of the crystal bulk as well), while such an effect is not visible in the case of the sample 3-300-3, prepared from the very same wafer, meaning that the chemical differences or defect concentration could hardly play any role. The sample 3-300-6 can also be characterized as imploded, while sample 3-300-2 demonstrates only several spike domains inside the poled domain, and thus presents the early stage of a DW implosion. Panels (c, f, i, and l) show the resulting I-V characteristics. The substantial differences in the geometrical structure are tightly intertwined with significant differences in the I-V curves; the current levels between samples can vary up to two orders of magnitude; the measure of 'non-linearity' from sample to sample is also quite different, with sample 3-300-6 (k) demonstrating the least 'rectifying' behavior.}
\end{figure*}

When comparing the current-voltage characteristics before and after the conductivity enhancement process among samples from the same batch, such as six samples from wafer~2 (samples 2-200-1...6) and six samples from wafer~3 (samples 3-300-1...6), a significant difference in pre- and post-enhancement conductance is evident, see Fig.~\ref{fig:Current_Change}. We can note that a clear current enhancement was observed in all cases, with the maximal factor of 1.9$\times$10$^{7}$ been reached in the case of sample 3-300-6 at +10~V; the minimal enhancement factor at +10 V is 15 as obtained for sample 3-300-3. Both of them are outliers in terms of their magnitude; the average enhancement factors are 1.7$\times$10$^{5}$ for -10~V and 2.5$\times$10$^{6}$ for $+$10~V.
Despite their identical preparation and uniform conductivity enhancement procedures, the conductance of these samples varies notably, sometimes even by multiple orders of magnitude. The SHGM images obtained \emph{after} the enhancement process explain these conductance disparities to some extent, since they unveil corresponding variations in the geometrical structures of the domain walls. Some samples still display an "imploded" state [e.g. sample 03-300-6, see Fig.~\ref{fig:SHG_post-enhancement_selection}(j, k)], even though the voltage ramp was halted after reaching a current value of 10$^{-6}$A. Conversely, other samples exhibit random, highly irregular structures that differ significantly from each other [see Fig.~\ref{fig:SHG_post-enhancement_selection} (d, g, and e)] and can not be categorized as being imploded. Notably, the current-voltage relationship during the conductivity enhancement procedure varies noticeably from the outset. This is evident from the voltage and time dependencies of the current during the procedure (Fig.~S7(a) and (d) of the supplementary material). For instance, the current of samples 2-200-3 and 2-200-4 differ by two orders of magnitude at a low voltage of 25~V during enhancement already. Similarly, the maximal stabilization voltages can span from -308~V to -478~V (Fig.~S7(d) in the supplementary material). The different DW geometry of the given samples is accompanied by a substantial difference in the resulting \emph{post-enhancement} I-V characteristics, see Fig.~\ref{fig:SHG_post-enhancement_selection}(c, f, i, and l). Here we see, how the maximal current differs by up to three orders of magnitude between investigated samples; the linearity of the curves is also quite different, with imploded sample 3-300-6 [Fig. \ref{fig:SHG_post-enhancement_selection}(l)] being an outlier in terms of 'linearity'. Furthermore, the analysis of the rest of the samples considered here does not allow us to extract an unambiguous correlation between the geometrical structure of the domain wall, in particular its average inclination towards the z-axis, and its conductive properties yet; the differences, so far, appear to be random. These results point us toward potential differences arising from subtle dissimilarities during the poling procedures or inherent variations within the crystal itself, including the distribution of crystalline defects and impurities within both the bulk of the crystal and its surface, as well as the individual electronic defect structure of the DW-electrode junction.

%\newpage

%\newpage
The reasons behind the limited reproducibility in CDW production remains to be studied even more profoundly. It is well understood that the shape of inclined domain walls directly impacts their conductance \cite{Wolba2018}, which is linked to the sin($\alpha$) where $\alpha$ represents the angle of the DW's inclination with respect to the crystal's z-axis. This inclination is sensitive to surrounding electric fields, including external ones \cite{Lu2019, Godau2017} and internal fields within the crystal, like depolarization and defect fields \cite{Tian2008, Carville2016}. The influence of external fields during DWC enhancement is still not thoroughly studied. The transformation of a hexagonal structure into a rounded triangle due to DWs contracting inwardly, except at 3 fixed points along the triple y-axis symmetry, as described by Kirbus \emph{et al.}~\cite{KGWB2019} and Godau \emph{et al.}~\cite{Godau2017}, seems more an exception than a rule after this large scale study.

For instance, comparing the conductance of samples from two different wafers --  wafer~2 in Fig.~\ref{fig:Current_Change} with a thickness of 200~\textmu m, and wafer~3 in Fig.~\ref{fig:Current_Change} with a thickness of 300~\textmu m - reveals that CDWs in the thicker wafer tend to exhibit lower conductance (three out of six samples have a maximal current below 0.1~\textmu A; while for the thinner wafer, the maximum currents for all samples were around 1~\textmu A). These differences could be explained by (i) slight variations in the chemical composition of the crystals and (ii) by potential differences in the electric field gradient during the enhancement procedure. For both batches the voltage ramp-up was the same, resulting in an electric field gradient 1.5 times higher for the thinner samples of batch 2.1. This suggests that the electric field gradient might be directly connected to the final conductivity of the structures. Hence, finding of optimal parameters for the procedure, including those that lead to domain implosion, becomes crucial as a future challenge.

Lastly, theoretical studies \cite{Sanna2017, Eliseev2009} highlight the high sensitivity of surface chemical and physical properties to the methods used in crystal production and the surrounding conditions. These aspects, in turn, influence geometry and thus the conductance of CDWs. Furthermore, experimental data from Kirbus~\emph{et~al.}~\cite{KGWB2019} indicates that the interface between CWD and metal electrode serves as the primary hindrance to charge transport due to the absence of DW inclination near the surface. This underscores the need for a more comprehensive investigation into the interface conditions between the crystal and electrode, e.g., by varying the contact metal and the surface preparation protocol. 

\section{Summary and Outlook}

In this work, a method for reproducible fabrication of hexagonal ferroelectric domain structures into 200~\textmu m and 300~\textmu m-thick 5-mol$\%$ MgO-doped LiNbO$_3$ single crystals was presented. We used a 325 nm UV-light-assisted electric-field poling setup and studied a variety of parameters influencing the fabrication. From these, key parameters such as electric field strength and poling length were identified and controlled by a poling protocol. The resulting domain structures have been analyzed with polarization-sensitive as well as second-harmonic generation microscopy and evaluated with respect to domain area and uniformity. The reproducibility of the protocol was tested for two different poling pulse lengths, and upon achieving this, the dependency of domain area on poling pulse duration for different wafers and electric field strength was measured.

This connection varies among different wafers, even when all other parameters remain constant, indicating high sensitivity to material composition and defect density.
Overall, the inclination of the freshly poled domains is small and therefore the resulting DWs have low conductance.
To increase the average inclination, the conductance enhancement procedure of Godau \emph{et al.}~\cite{Godau2017} was applied to the domain structures. Initially, the voltage was steadily increased by 4~V/s to 500~V. All domains undergoing this protocol "imploded", i.e. disintegrated into many needle-like subdomains, once the current exceeded a maximum of 10$^{-5}$~A.
To prevent this implosion in the ensuing measurements, a maximal current of 10$^{-6}$ was implemented and the voltage ramp was stopped upon reaching this value, leading to a current-control rather than voltage-control scheme. 

The testing of this improved procedure has revealed that despite the identical conditions during the conductivity enhancement process, both the geometry of the domain wall structures and their conductance - a parameter closely linked to the latter factor - still exhibit significant variations, although only a few show signs of implosion. For instance, the difference in the maximal conductance among the samples can vary by up to three orders of magnitude. Likewise, the geometrical structure of the domain walls still displays a certain diversity, including pinning sites, loss of symmetry in relation to the crystal structure, and even implosion of the domains.

The breadth of the obtained data emphasizes the necessity for further investigation and standardization of the DWC enhancement process, as well as research on the correlation between conductance and geometric changes throughout the procedure. In this regard, the implementation of \emph{in-situ} SHG during the enhancement process shows promise\cite{Kirbus2019}. Additionally, a more comprehensive examination of the influence of intrinsic electric fields, particularly those on the surface of the lithium niobate crystal \cite{KGWB2019}, is essential. Such fields can be affected by various factors, including the electrode material, surface preparation and processing method, and chemical composition of the crystal -- their systematic study will be subject to future research.

The results of this work are far from conclusive, as the domain growth as well as the DWC enhancement procedure are highly dynamical processes sensitive to many local and nanoscopic influences, which do not easily succumb to a straightforward macroscopic description and modelling. On the contrary, this extensive work highlights the fact that besides application based research on memories, neural networks, or rewritable electronics, more fundamental studies on the interplay and dynamics of domains and domain walls with localized defects is mandatory, if ever the dream of true DW-based nanoscopic electronics should become real.

\section*{Supplementary Material}

See the supplementary material for a complete list of all samples; in-depth information on domain (wall) imaging by PLM and SHGM including the mathematical extraction of the correct domain area; details on the domain growth process during UV-assisted poling with particular focus on the development of the DW inclination; results of preliminary experiments investigating the domain area's dependence on laser intensity and NaCl concentration during UV-assisted liquid-electrode poling; diagrams with the complete I-V characteristics of sample batches 2.1 and 3.1 \textit{during} and \textit{after} DWC enhancement.

\section*{Acknowledgements}

We acknowledge financial support by the Deutsche Forschungsgemeinschaft (DFG) through joint DFG--ANR project TOPELEC (EN~434/41-1 and ANR-18-CE92-0052-1), the CRC~1415 (ID: 417590517), the FOR~5044 (ID:
426703838; \url{https://www.for5044.de}), as well as through the Dresden- W\"urzburg Cluster of Excellence on "Complexity and Topology in Quantum Matter" - ct.qmat (EXC 2147, ID: 39085490). This work was supported by the Light Microscopy Facility, a Core Facility of  the CMCB Technology Platform at TU Dresden. I.K.'s contribution to this project is also co-funded by the European Union and co-financed from tax revenues on the basis of the budget adopted by the Saxon State Parliament. Furthermore, we thank Thomas Gemming and Dina Bieberstein for assistance in wafer dicing.

\section*{Author Declarations}

\subsection*{Conflict of Interest}
The authors have no conflicts to disclose.

\subsection*{Author Contributions}
\textbf{Julius Ratzenberger:} Data curation (equal); Formal analysis (equal); Investigation (equal); Validation (equal); Visualization (equal); Writing -- original draft (equal); Writing -- review \& editing (supporting). \textbf{Iuliia Kiseleva:} Data curation (equal); Formal analysis (equal); Investigation (equal); Validation (equal); Visualization (equal); Writing -- original draft (equal); Writing -- review \& editing (supporting). \textbf{Boris Koppitz:} Investigation (supporting); Visualization (supporting). \textbf{Elke Beyreuther:} Funding acquisition (supporting); Methodology (supporting); Supervision (equal); Writing -- review \& editing (lead). \textbf{Manuel Zahn:} Formal analysis (supporting); Software (lead). \textbf{Joshua Gössel:} Investigation (supporting). \textbf{Peter A. Hegarty:} Investigation (supporting); Visualization (supporting). \textbf{Zeeshan H. Amber:} Investigation (supporting); Visualization (supporting). \textbf{Michael Rüsing:} Conceptualization (lead); Funding acquisition (equal); Methodology (lead); Project administration (equal); Supervision (equal); Writing -- review \& editing (supporting). \textbf{Lukas M. Eng:} Conceptualization (equal); Funding acquisition (lead); Methodology (supporting); Project administration (equal); Resources (lead); Supervision (equal); Writing -- review \& editing (supporting).

\section*{Data Availability}

The data that support the findings of this study are available from the corresponding author upon reasonable request.

\section*{References}

\bibliography{literature}

\clearpage
\newpage

\beginsupplement

\onecolumngrid

\section*{Supplementary Material}

\subsection{Table of all samples}

\begin{table}[h]
}                                                                                                                                                                                                                                                                                                                                                                                                                                                         
                                                                                                                                                                                 & 3-300-07                                             \\ \cline{6-6} 
 
&                                                 &                                                                                                                   &                                                                                                                                                                                                                                             &                                                                                                                                                                                                                                     

& 3-300-08                                             \\ \cline{6-6} 
                                                                                                                                                                                &                                                 &                                                                                                                   &                                                                                                                                                                                                                                             &                                                                                                                                                                                                                                      & 3-300-09                                             \\ \cline{6-6} 
                                                                                                                                                                                &                                                 &                                                                                                                   &                                                                                                                                                                                                                                             &                                                                                                                                                                                                                                      & 3-300-10                                             \\ \cline{6-6} 
                                                                                                                                                                                &                                                 &                                                                                                                   &                                                                                                                                                                                                                                             &                                                                                                                                                                                                                                      & 3-300-11                                             \\ \cline{6-6} 
                                                                                                                                                                                &                                                 &                                                                                                                   &                                                                                                                                                                                                                                             &                                                                                                                                                                                                                                      & 3-300-12                                             \\ \cline{6-6} 
                                                                                                                                                                                &                                                 &                                                                                                                   &                                                                                                                                                                                                                                             &                                                                                                                                                                                                                                      & 3-300-13                                             \\ \hline
\end{tabular}
\end{table}

%\clearpage

\subsection{Imaging of domains and domain walls by polarization-sensitive microscopy (PLM) and second-harmoic generation microscopy (SHGM)}

To achieve reproducibility in domain- and domain wall fabrication, domains of equal size are a first prerequisite. To control the results of domain engineering, the size and structure of each domain are determined by two independent imaging techniques, a cross-polarization light microscope (PLM) and a second harmonic generation microscope (SHGM). An example result obtained for the same domain for each imaging technique is shown in Figs.~2(a) and (b) of the main text. Both methods see widespread use for imaging of domains \cite{Kaempfe2014, KGWB2019, Kaneshiro2008, Cherifi-Hertel2017, Otko1993} and have their own advantages. In particular, PLM has a very long history for imaging of domains due to its fast and easy application, but only offers a 2D projection of the domain structure. In contrast, SHGM allows for true 3D reconstruction with voxel sizes down to $< 1$\textmu m$^3$, but with more experimental effort. Therefore, we are systematically comparing imaging results obtained for both methods.

The contrast in PLM relies on a change of local birefringence in the domain wall and its surrounding area due to high internal stress and an incomplete screening due to reversed polarization. Therefore, the material near a DW shows a different optical anisotropy than a bulk crystal. Consequently, due to the electro-optic effect, the linearly polarized light passing close to the DW will experience a different phase shift and rotation of the light's polarization~\cite{Soergel2005}. The unstressed crystal is equally birefringent independent of domain orientation and, therefore, the z- and z+ regions are indistinguishable. We therefore filter out all light with the same polarization using a second linear polarizer that is rotated by 90$\symbol{23}$. In the resulting image, as depicted in Fig.~2(a) of the main text, only the area around the domain wall structure itself exhibitss a contrast. To achieve good reproducibility, we positioned the sample with the z+ surface facing upwards and the microscope is focused on this surface. The spike domain around the nucleation center is not visible, as the structure is not continuous and too small to create a sufficiently large phase shift.

The contrast mechanism in SHGM relies on the local changes of the second-order nonlinear optical response at the domain wall. Most commonly, due to the so-called Cerenkov-type phase matching, the signal at domain walls is enhanced compared to its surroundings \cite{Hegarty2022}, which is the dominating effect in this work. The experimental setup for SHGM used here consists of a \emph{Zeiss LSM980} microscope combined with a fs-pulsed laser system \emph{SpectraPhysics InSight X3} (690–1300~nm, 3.5~W, 120~fs pulse width), which was tuned to 900~nm wavelength. Typically, average powers in the range of 10~mW were required to generate a signal. For all measurements, an appropriate objective lens with an NA of 0.8 was used providing approximately 0.5~\textmu m lateral and 4~\textmu m axial resolution within the material. As SHGM is a scanning technique, the beam is scanned via a galvano-mirror system providing a 2D field of view of 400$\times$500~\textmu m$^2$. For 3D analysis or to image larger areas, the sample is mounted on a 3D positioning stage system. More details on the setup can be found elsewhere \cite{Reitzig2022, Hegarty2022}. The domain wall itself is visible as a strong bright maximum in the SHGM images, as shown in  Fig.~2(b) of the main text. In the exemplary scan, there are two main features visible: An almost regular hexagonal shape confining the main poled domain area. In the middle of this area a tiny, irregular domain shape appears. These inner irregular domains typically only appear near the z+ surface and often vanish completely close to the z- surface. We identify these structures as spike domains, which form due to spontaneous back-poling at the nucleation center during the poling process~\cite{Kim2001}. As these are not connecting through the sample, they can usually be ignored for electrical measurements. Stacked images from different depths can be merged together to form a 3D projection of a domain wall as shown in Fig.~S4(a) of the supplementary material, where we omit the spike domain for better clarity. A key feature of SHGM is that it allows mapping the DW inclination angle with respect to the polar z-axis, which is directly connected to the number of screening charges present at the DW and thus the DWC~\cite{Kirbus2019}.

As shown exemplarily in Fig.~2 of the main text, the undisturbed poled domains in MgO:LN from our setup always form a hexagon with a constant internal angle of approx. 120$\symbol{23}$. This configuration is energetically most favorable in MgO:LN and has its origin in the specific crystal structure of LN~\cite{Shur2006}. Small deviations of the internal angle arise due to pinning of the domain wall during growth, but are below 2$\symbol{23}$. During our experiments, the hexagons are usually not perfectly uniform, i.e., the length of each side can vary substantially and can not be assumed equal. Therefore we gauge the domains not by the diameter but by calculating the overall domain area for both methods. A detailed description of this measurement scheme is given in the supplementary material (Secs.~C and D).  

As we use results from both SHGM and PLM, a comparison is obvious and reveals significant differences: Only in the SHGM we can observe the real thin domain wall within a defined depth. In the PLM a thicker area affected by the strain-induced difference in anisotropy is visible because a minimum optical phase shift is required to visualize the domain structure. Moreover, with SHGM we can selectively "look" at any given depth within the sample, while in the PLM we obtain a 2D projection of the domain structure only. In direct comparison, the imaging method PLM leads to a constant overestimation of the domain size  for domains larger than 3$\times10^{4}$~\textmu m$^{2}$ of around 5~$\%$. Therefore, for our studies we rely on the SHGM data, however, due to this linear error between both methods, similar conclusions can be drawn with both methods for domains larger than 3$\times10^{4}$\textmu m$^{2}$. A more detailed discussion of the comparison between both methods can be found in the following section.

\subsection{Correlation of the domain sizes determined via SHGM and PLM}

\begin{figure*}[!htb]
\includegraphics[width=0.9\textwidth]{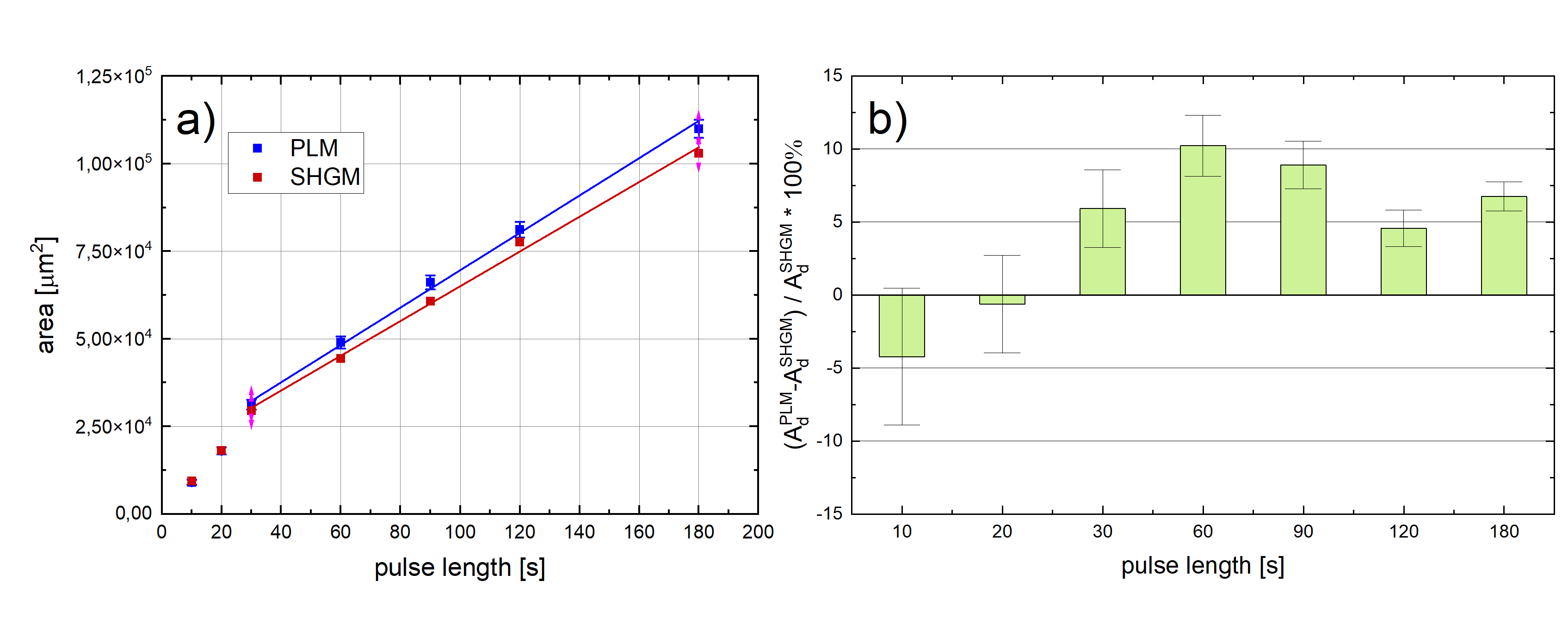}
\caption{\label{fig:SHGvsPLM_detailed} Comparison between the two imaging methods PLM and SHGM. In absolute values (a), the difference for smaller domains below 3$\times10^{4}$\textmu m$^{2}$ is not noticeable and increase linear with longer pulse lengths. A more detailed view into the relative differences (b) reveals -- in a smaller area -- an underestimation of the area up to -5~$\%$ of the PLM methods in comparison to the SHGM. However, no definitive conclusion can be made because the error range is larger than the difference. The overestimation of the area for larger domains by PLM already visible in (a) is further confirmed.}
\end{figure*}

We have used two different imaging methods in this work, SHGM and PLM. To check our results with both methods, we compared them over a series of poling pulse lengths $t_{p}$, as shown in Fig.~\ref{fig:SHGvsPLM_detailed}. 
For small areas produced by a $t_{p}$ of 10 to 20~seconds, we found good agreement between the SHGM and PLM images. A more detailed view is provided by Fig.~\ref{fig:SHGvsPLM_detailed}(b), which shows the difference between the methods relative to the SHGM area. In a smaller area, the PLM method appears to underestimate the area up to -5$\%$, but no definitive conclusion can be made because the error range is larger than the difference. For larger ranges above 3$times10^{4}$\textmu m$^{2}$, a clear trend toward an overestimation of the domain area with the PLM method is evident, with an average of 7.2$\%$ larger areas in the PLM images. 
A combination of factors contributes to this effect. The PLM method overestimates the area of a domain: a real domain in MgO:LN has slightly rounded corners, while our theoretical model assumes sharp corners, as described in the supplementary material (Sec.~D). 
This leads to somewhat larger calculated domain sizes. Additionally, the domain wall itself is not visible in the PLM, only the optical effects close to it. As can be seen in Fig. \ref{fig:SHGvsPLM_detailed}(a), the boundary region between z- and z+ domains is wider than the width of the DW itself. As explained before, the DW itself is not visible, only the birefringence due to stress or incomplete screening near the wall. Therefore a small offset between measured and actual position exists. To keep this offset consistent, we always measured at the outer dark edge of the domain boundary, which likely contributes to a further overestimation of the area. 
These two reasons of overestimation of the domain area do not exist in the SHGM method: to extract the domain area from the SHG image, we use a Python script to count the pixels within the domain area and convert them to the area in \textmu m$^{2}$. Thus, we circumvent the overestimation due to the theoretical model. In addition, the SHGM scans map the domain wall itself and therefore do not suffer from the same "real"-domain-wall problem of the PLM method. 
The areas of small domains were probably underestimated because it was difficult to measure them accurately in the PLM. Additionally, the effect of rounded corners contributes stronger in the case of small domains.

\subsection{Determination of the area of a hexagon with equal internal angles}

To calculate all distances between different corners and the area of the hexagonal domain structure, it is sufficient to know the three heights of the hexagonal shape as well as one side length. While it would also be possible, to use certain other combinations of four different lengths, e.g., four side lengths, the uncertainty is much smaller for the former combination, thus making it the most feasible approach.

The following calculations are made under the assumption that the angles between adjacent sides of the hexagon are all equal to $\alpha=2 \pi/3$ (120$\symbol{23}$), which, due to symmetry, is a plausible requirement.

\begin{figure}[!htb]
\includegraphics[width=\textwidth]{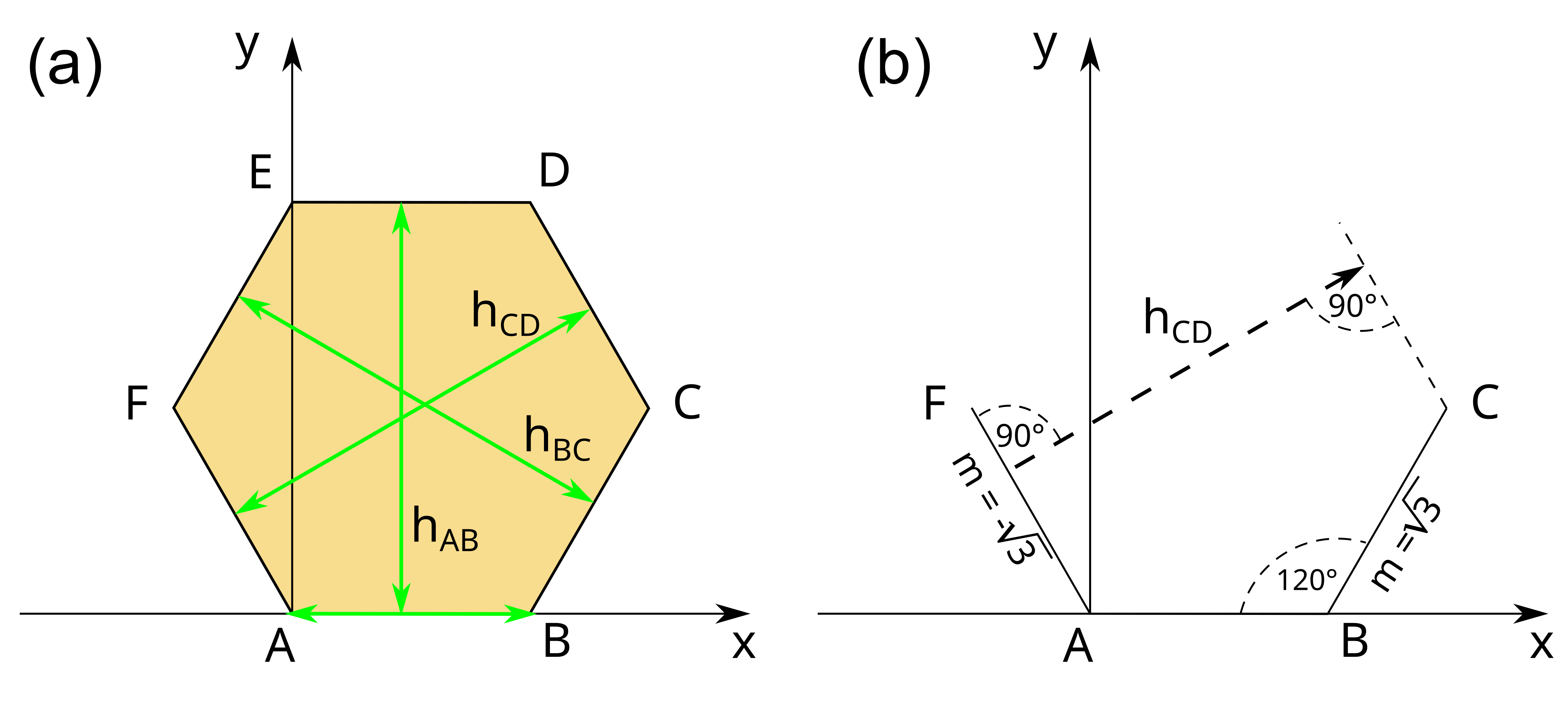}
\caption{(a) Visualization of the hexagon with labeling of the corners. Corner A lies directly in the coordinate origin and the green lines show the measured distances. (b) Parallel shift of line $AF$ with the slope $m=-\sqrt{3} $ by $h_{DC} $ to calculate line $CD$. }
\label{fig:hex_all}
\end{figure}

We define a coordinate system with the bottom left corner of the hexagon at its origin and the bottom side on the $x$ axis, as sketched in Fig.~\ref{fig:hex_all}(a).
The six different corners of the hexagon are labeled counterclockwise, starting at the origin with point $A$ at coordinates $(x_A, y_A)=(0,0)$, followed by the bottom right point $B$ at $(x_B,y_B)=(l_{AB},0)$, the middle right point $C$ at $(x_C,y_C)$, the top right point $D$ at $(x_D,y_D)=(l_{AB},h_{AB})$, the top left point $E$ at $(x_E,y_E)=(0,h_{AB})$ and the middle left point $F$ at $(x_F,y_F)$, where the heights $h_{AB}$, $h_{BC}$ and $h_{CD}$ as well as the length $l_{AB}$ are known, as shown in green in Fig.~\ref{fig:hex_all}(a).
To calculate the remaining unknown coordinates $x_C, y_C, x_F$, and $y_F$, we first note that $C$ is the intersection of the lines $BC$ and $CD$.
The line $BC$ has a slope of $m_{BC}=\Delta y/\Delta x=\tan(\pi-\alpha)=-\tan(\alpha)=-\tan(2\pi/3)=\sqrt{3}$, while the slope of the line $DC$ is $m_{CD}=-\sqrt{3}$. 
We thus write the equations of the two lines as:
\begin{align}
BC:&&y_{BC}(x)&=\sqrt{3}(x-l_{AB})\\
CD:&&y_{CD}(x)&=-\sqrt{3}x + 2 h_{CD},
\end{align}
where the second line, $CD$, can be constructed by parallel translation of the line $AF$, which has the same slope, but an intercept that is shifted according to Fig.~\ref{fig:hex_all}(b).
Solving $y_{BC}(x_C)=y_{CD}(x_C)$ for $x_C$ and substituting this back into one of the two equations in order to get $y_C$ leads to $(x_C, y_C)=(l_{AB}/2+h_{CD}/\sqrt{3}, -\frac{\sqrt{3}}{2}l_{AB}+h_{CD})$.

Analogously, we can derive the equations:
\begin{align}
AF:&&y_{AF}(x)&=-\sqrt{3}x\\
EF:&&y_{EF}(x)&=\sqrt{3}(x-l_{AB}) + 2 h_{BC}
\end{align}
of the two lines $AF$ and $EF$, where $y_{EF}(x)$ results from a parallel translation of the line $BC$. The intersection $(x_F,y_F)=(l_{AB}/2-h_{BC}/\sqrt{3}, -\frac{\sqrt{3}}{2}l_{AB}+h_{BC})$ yields the coordinates of the point $F$.

The points $D$ and $E$ can now be calculated from the intersection of $CD$  and $EF$ with the line
\begin{align}
DE:&&y_{DE}(x)&=h_{AB},
\end{align}
respectively, which leads to $(x_D,y_D)=((2h_{BC}-h_{AB})/\sqrt{3},h_{AB})$ and $(x_E,y_E)=(l_{AB}+(h_{AB}-2h_{BC}/\sqrt{3}, h_{AB})$.
A summary of the coordinates can be found in Tab.~\ref{tab:hexagon_coords}.

The uncertainties of the coordinates are calculated via error propagation assuming independent variables as
\begin{align}
\Delta x_i &= \sqrt{\left(\frac{\partial x_i}{\partial l_{AB}}\right)^2 (\Delta l_{AB})^2 +
\left(\frac{\partial x_i}{\partial h_{AB}}\right)^2 (\Delta h_{AB})^2+
\left(\frac{\partial x_i}{\partial h_{BC}}\right)^2 (\Delta h_{BC})^2 } \text{...}  &\\
&\text{...}\overline{+\left(\frac{\partial x_i}{\partial h_{CD}}\right)^2 (\Delta h_{CD})^2 } \notag\\
& ~&~& \notag \\
\Delta y_i &= \sqrt{\left(\frac{\partial y_i}{\partial l_{AB}}\right)^2 (\Delta l_{AB})^2 +
\left(\frac{\partial y_i}{\partial h_{AB}}\right)^2 (\Delta h_{AB})^2+
\left(\frac{\partial y_i}{\partial h_{BC}}\right)^2 (\Delta h_{BC})^2 } \text{...}  & \\
&\text{...}\overline{+\left(\frac{\partial y_i}{\partial h_{CD}}\right)^2 (\Delta h_{CD})^2 }\notag,
\end{align}
where $i\in \lbrace A,B,C,D,E,F\rbrace$. Assuming that all four lengths $l_{AB}$, $h_{AB}$, $h_{BC}$, and $h_{CD}$ have the same error $\Delta$, we obtain the uncertainties shown in Tab.~\ref{tab:hexagon_coords}

\begin{table}
\renewcommand{\arraystretch}{2} 
\begin{tabular}{l|cccccc}
\hline\hline
point           & $A$ & $B$     & $C$                                       & $D$                               & $E$                                      & $F$ \\ \hline\hline
$x$ coordinate  & 0   & $l_{AB}$& $\frac{l_{AB}}{2}+\frac{h_{CD}}{\sqrt{3}}$& $\frac{2h_{BC}-h_{AB}}{\sqrt{3}}$ & $l_{AB}+\frac{h_{AB}-2h_{BC}}{\sqrt{3}}$ & $\frac{l_{AB}}{2}-\frac{h_{BC}}{\sqrt{3}}$ \\ \hline 
$y$ coordinate  & 0   & 0       &  $-\frac{\sqrt{3}}{2}l_{AB}+h_{CD}$       & $h_{AB}$                          & $h_{AB}$                                 &  $-\frac{\sqrt{3}}{2}l_{AB}+h_{BC}$ \\ \hline 
$x$ uncertainty & 0  & $\Delta$ & $\sqrt{\frac{7}{12}}\Delta $              & $\sqrt{\frac{5}{3}}\Delta$        & $\sqrt{\frac{8}{3}}\Delta $              & $\sqrt{\frac{7}{12}}\Delta $ \\ \hline 
$y$ uncertainty & 0  & 0        &  $\sqrt{\frac{7}{4}}\Delta $              & $\Delta$                          & $\Delta$                      & $\sqrt{\frac{7}{4}}\Delta $ \\
\hline\hline
\end{tabular}
\caption{Summary of the coordinates for the six corners of the hexagon, calculated from the three heights and one side length. The last two lines show the uncertainties assuming that all four lengths $l_{AB}$, $h_{AB}$, $h_{BC}$, and $h_{CD}$ exhibit the same error $\Delta$.}
\label{tab:hexagon_coords}
\end{table}

For every pair $(i,j)$ of corners ($i, j \in \lbrace A,B,C,D,E,F\rbrace$), the distance $s_{i,j}$ between these corners, i.e., a side lengths of the hexagon or one of the diagonals, can be calculated using the Pythagorean theorem via
\begin{equation}
s_{i,j}= \sqrt{(x_i-x_j)^2+(y_i-y_j)^2}.
\end{equation}
The uncertainty $\Delta s_{i,j}$ is given by
\begin{equation}
\Delta s_{i,j}= \frac{\sqrt{(x_i-x_j)^2(x_i^2(\Delta x_i)^2+ x_j^2 (\Delta x_j)^2)+(y_i-y_j)^2(y_i^2(\Delta y)_i^2+ y_j^2 (\Delta y)_j^2)}}{s_{i,j}}.
\end{equation}

The area of the hexagon can be calculated via the shoelace formula (also known as Gauss' area formula), which states that the area $A$ of a polygon with corners $i\in \lbrace 1,\dots, n\rbrace$  can be calculated via
\begin{equation}
A=\frac{1}{2} \sum_{i=1}^{n} (y_i+y_{i+1})(x_i-x_{i+1}),
\end{equation}
which can be applied to our hexagon by setting $i=1 \leftrightarrow A$, $i=2 \leftrightarrow B$, $i=3 \leftrightarrow C$, $i=4 \leftrightarrow D$, $i=5 \leftrightarrow E$, and $i=6 \leftrightarrow F$. 

\newpage

\subsection{Comparison between domain area, hexagon height and side lengths}

\begin{figure*}[!htb]\includegraphics[width=\textwidth]{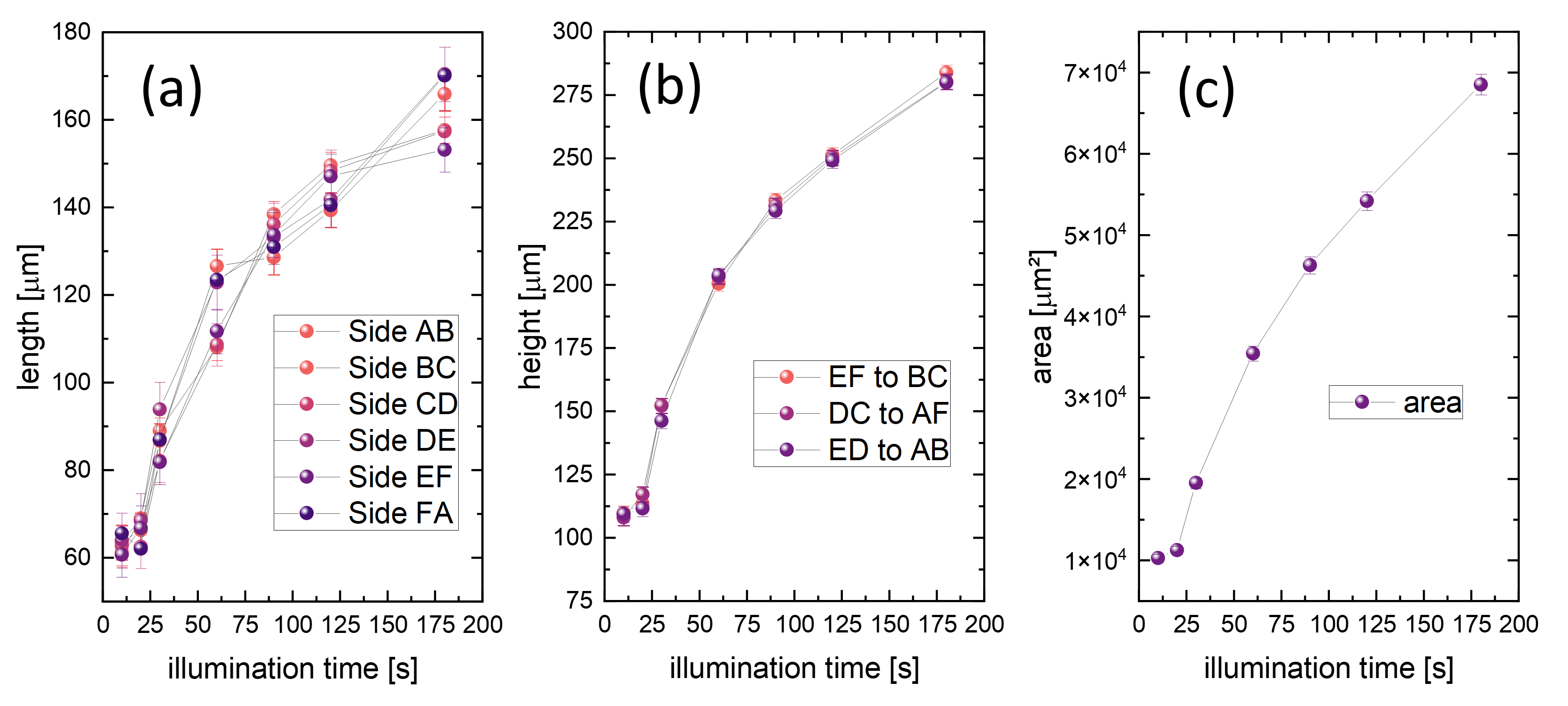}
\caption{\label{fig:heightseize900old} (a)-(c) Measurement of the three heights, the length of each side and the total area of the domains of batch 1.3 with an increasing $t_{p}$. All metrics show the same overall trend. A larger scattering is shown by the length of the individual sides. This is directly related to the influence of disturbances and pinning during growth. Nevertheless the same trend as for the total domain area is recognizable.}
\end{figure*}

In order to calculate the domain area, as described in Sec.~D, we measured the three heights as well as one side length of the hexagon. These lengths show the same trend as the overall domain area, but with a larger fluctuation, as pictured in Fig.~\ref{fig:heightseize900old}. We therefore use the overall domain area as a more robust metric to evaluate the result of the domain growth process.

%\newpage

\subsection{Development of the domain-wall inclination during the UV-assisted poling process monitored by SHG microscopy}

\begin{figure*}[!htb]
\includegraphics[width=\textwidth]{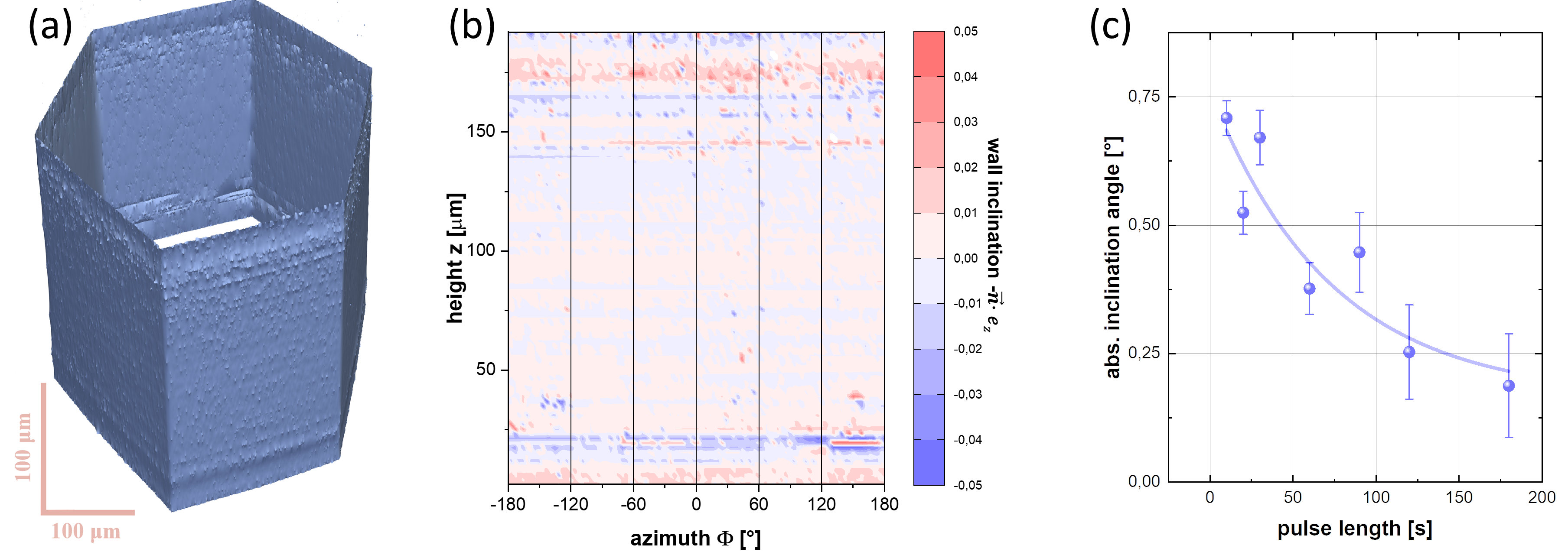}
\caption{\label{fig:Inclination} (a) A 3D SHGM image of the poled hexagonal domain 2-200-07; the small spike domain in the center has been omitted for better visualization. (b) The inclination of a domain wall with respect to the z-axis is mapped to a cylindrical coordinate system. A blue coloring represent parts of the domain wall, where the polarization is directed from it (tail-to-tail), which result a negative tilt angle and parts where the polarization is directed toward the domain wall (head-to-head), which have a positive tilt angle, are colored in red. (c) Average tilt of the domain walls with respect to the pulse length $t_{p}$. The inclination flattens out with longer poling pulse durations, as indicated by the blue line.}
\end{figure*}

Domains typically start as spike domains with an initial inclination with respect to the z-axis due to the role of domain wall conductivity in providing initial screening until the domain has reached the other side. The domains should flatten out, once they have fully penetrated the crystal \cite{Sturman2022}. To test this hypothesis, we investigated samples from batch 2.2, with different growth times. The 3D data of the domain wall provides an insight into the inclination of the domain wall with respect to the z-axis, as pictured in Fig.~\ref{fig:Inclination}(b). Here we unfolded the six hexagon surfaces on a 2D map to plot the average inclination with respect to the z-axis. The average inclination is uniformly small and only slightly increases close to the surface. These small angles do not cause increased conductivity compared to the bulk crystal at the domain walls. In Fig.~\ref{fig:Inclination}(c) we calculated these inclination angle averages for an increasing range of poling pulse lengths. The average inclination angle clearly decreases with the growth time as the domain area increases. 
This observation is consistent with the theory of nucleation and growth of domains in LN. The irradiation with a sub-bandgap laser source leads to an influx of charge carriers in the irradiated area. \cite{Wengler2005, Wang2009}. A nucleation center forms in this area around a defect or impurity, as these are the local minima in the coercive field. Subsequently, a spike domain begins to grow around the nucleation center. These structures do not initially reach the z-surface because they have a large tilt angle with respect to the z-axis and thus resemble a cone.
This spike domain then extends mainly downward through the crystal until it reaches the -z-surface. At this point, the domain begins to grow only outward along the crystallographic y-axis -- as the screening charge can now be supplied via the outer electrodes -- forms its typical hexagonal shape, and the growth slows down as soon as the domain reaches areas that are not affected by the laser. The overall angle of inclination decreases with outward growth as the z-surface catches up to the same size as the z+surface \cite{Shur2005}, leading to a nominally flat domain wall with respect to the z-axis. 
During this expansion, kinks and disturbances in the DWs can form. They originate from additional defects or impurities and hinder the domain growth in a certain direction. This causes the domain wall to be "pinned" at a certain spot, until the potential difference to the rest of the continuously growing wall around is big enough to overcome it \cite{Lixin2003, Gopalan2007}. Most deviations from the perfectly regular hexagonal shape can be attributed to this phenomenon.

\subsection{Influence of the laser intensity during the UV-assisted liquid-electrode poling process on the domain size}

\begin{figure*}[!htb]
\includegraphics[width=0.5\textwidth]{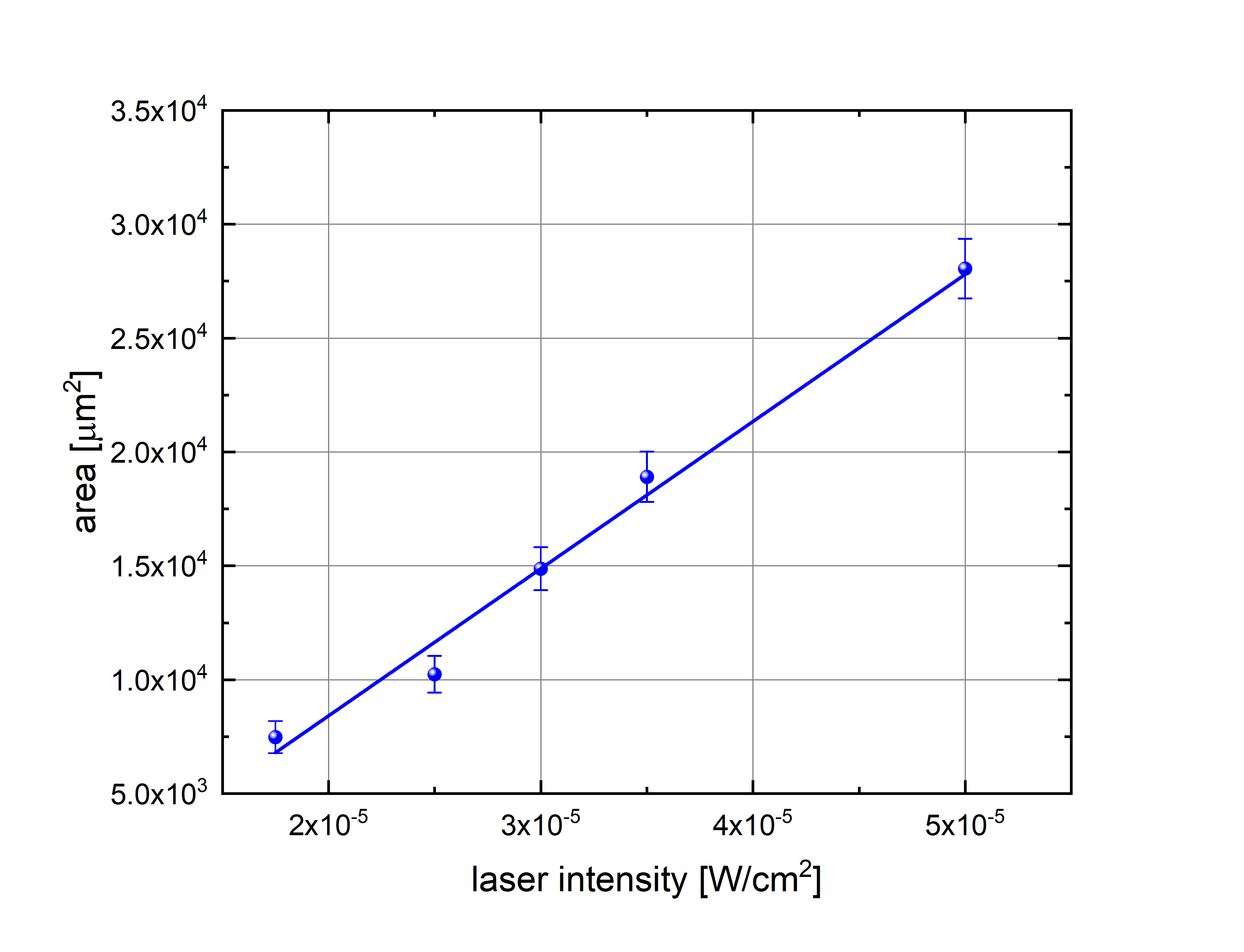}
\caption{\label{fig:tuning_graphs} 
Dependence of the domain area $A_d$ on the laser intensity. An approximately linear relationship is recognizable within the investigated intensity range.}
\end{figure*}

As a preliminary experiment, a range of laser intensities in the regions of 10$^{-3}$~W/cm$^{2}$ to 10$^{-7}$~W/cm$^{2}$ was tested· Too low intensities resulted in no or very slow domain growth, while a too high intensity lead to an almost immediately nucleation with an uncontrollable growth, resulting in very disturbed domain shapes and strong back-poling· We, therefore, chose an intensity range of 10$^{-5}$~W/cm$^{2}$, were a small nucleation center was observed, and a reproducible domain growth was possible. A study of this order of magnitude is displayed in Fig.~\ref{fig:tuning_graphs}. The domain area increases linearly with the intensity, as reported in the literature \cite{Sones2005,Sones2010}. Therefore, even an intensity fluctuation of 0.5~$\times$10$^{-5}$~W/cm$^{2}$ can yield a remarkable domain size difference. Hence, the laser intensity must be closely monitored. Thus, we measured its value at the start of each poling session and adjusted the intensity. For all measurements discussed in this work, the laser intensity on the sample was set to 2.8$\times$10$^{-5}$~W/cm$^{2}$.

\subsection{Influence of the NaCl concentration during the UV-assisted liquid-electrode poling process on the domain size}

\begin{figure*}[!htb]
\includegraphics[width=0.5\textwidth]{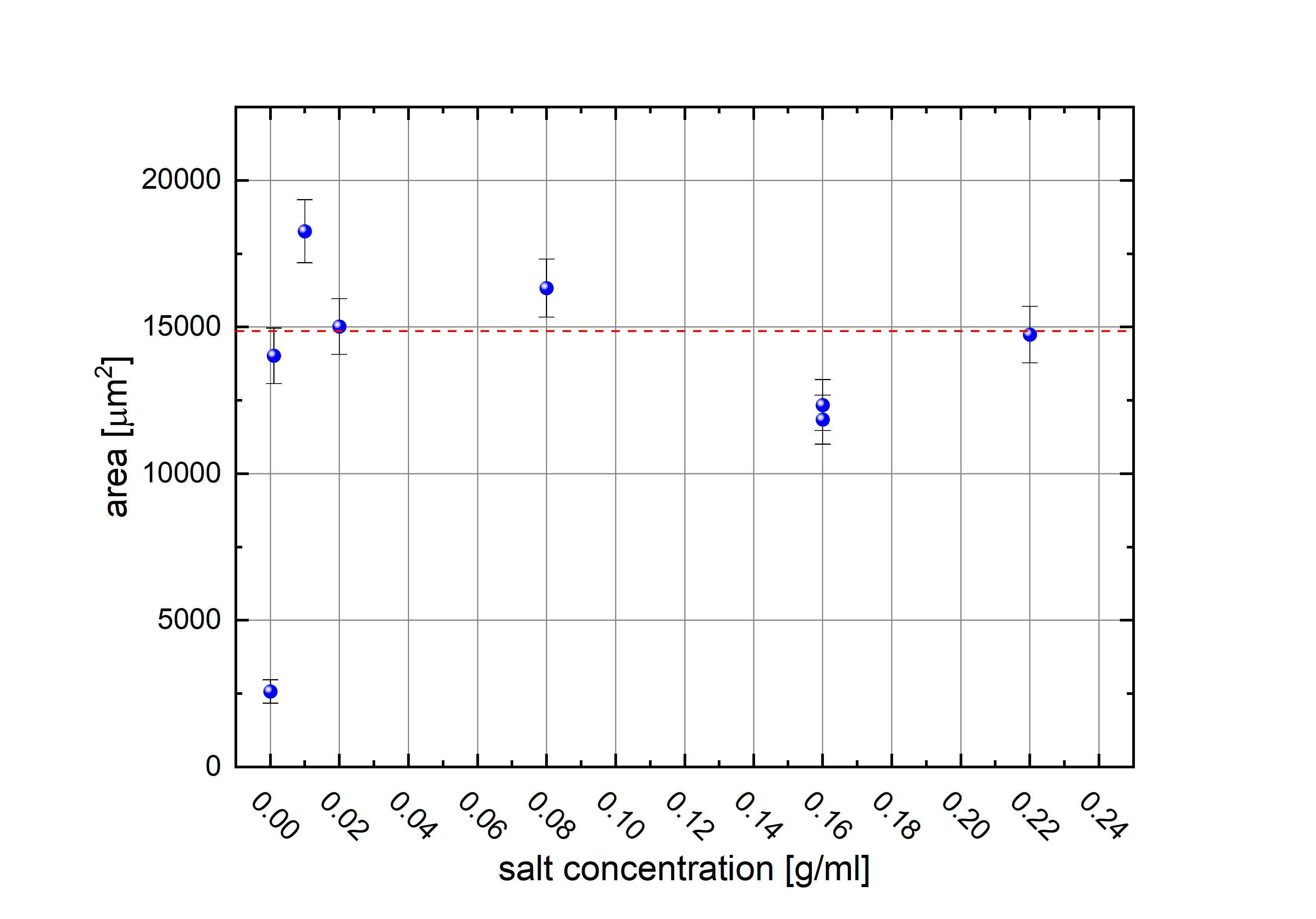}
\caption{\label{fig:salt_conc} Domain area as a function of the NaCl ("salt") concentration in the solution, which makes up the liquid electrode for domain engineering. The red line symbolizes the expected area for the poling parameters. Only in the case of pure deionized water, the domain growth is hampered. The expected area is achieved already with 0.1~g NaCl per 100~ml deionized water. Higher concentration did not lead to improvement/enlargement of the poled area.}
\end{figure*}

In early experiments sometimes huge deviations of the expected poling area under seemingly the same parameters (poling pulse length, electric field, laser intensity) were observed. To resolve this issue, the influence of the salt (here: NaCl) concentration in the liquid electrodes was investigated. We therefore systematically varied the salt concentration in a user-friendly range from 0 up to 20~g per 100~ml deionized water. Afterward, the same poling parameters of a $t_{p}$~=~30~s by an electric field of 4.0~kV/mm under a laser irradiation of 40 s were applied. 
In Fig. \ref{fig:salt_conc} the resulting domain area is plotted versus the NaCl concentration in g/ml. In the case of pure deionized water as liquid electrodes, the domain growth is hampered. As soon as 0.1~g of NaCl was dissolved in 100~ml of the liquid (deionized water), the
domain area normalized to the expected value. Further, no noticeable increase in domain area was observed with higher salt concentration.

It is well-known that the conductivity of a salt water solution increases linearly with salinity. Pure deionized water has a very low conductivity of 1~\textmu S/cm, while the conductivity of the 0.001~g/ml NaCl solution is already 3 orders of magnitude higher with 1~mS/cm. In the case of pure deionized water, the conductivity is too low to apply a uniform electric field to the sample surface, resulting in a poor poling result. The lowest tested salt concentration already reached a minimal conductivity, in which the electric field was uniformly applied to the sample. Higher conductivity of the liquid electrodes therefore did not lead to improvement of the domain growth.   
In addition, higher concentrations of salt resulted in residues in the liquid cell. These residues caused air bubbles and accumulated on the sample surface, both disturbing the growth of evenly hexagonal-formed domains. We therefore choose 2~g
NaCl per 100~ml deionized water, as it was an easy-to-use amount, while the residues in the cell and on the sample were still minimal.

\newpage

\subsection{Complete I-V characteristics of samples from batches 2.1 and 3.1 during and after DW conductivity enhancement}

\begin{figure*}[!htb]
\includegraphics[width=\columnwidth]{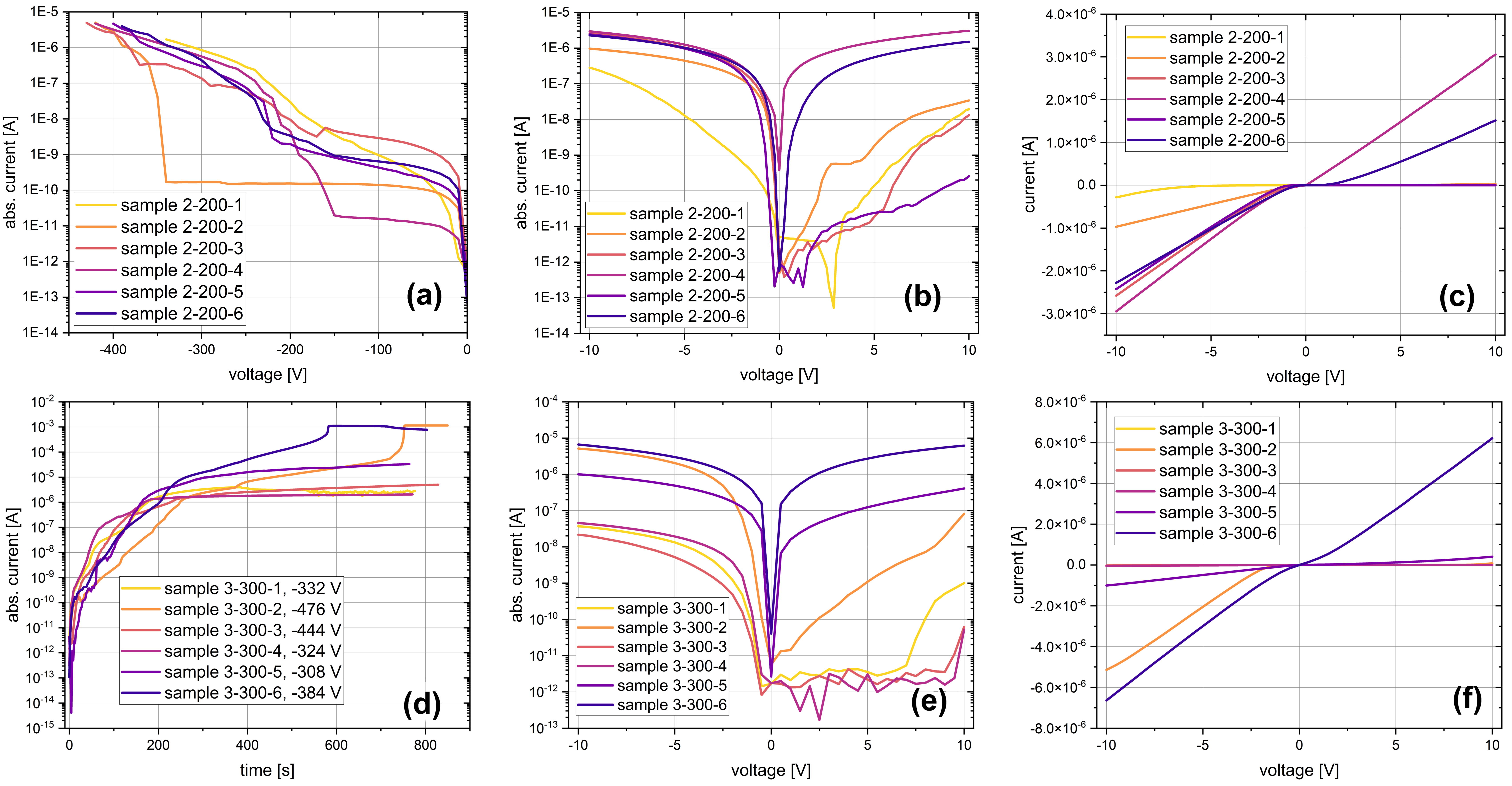}
\caption{\label{fig:IV_char_during_after_enh} (a, d) The change of the current during the conductivity enhancement procedure. For each sample the voltage was ramped up (4 V/s) until the current reached the value of 1 $\mu$A, then at the maximal voltage the sample was stabilized for 10 minutes; in picture (a) the change of the current with the voltage ramp-up is depicted; on picture (d), change of the current with time during both stages of procedure (voltage increase and stabilization time) is depicted; $V_{max}$ is indicated near the name of the sample. The difference in conductance between the samples during the DWC enhancement procedure can be observed from the very beginning of the procedure. As a result, obtained I-V characteristics of the samples (b, c and e, f), namely, the maximal conductance between the samples, can differ from each other by up to five orders of magnitude. For convenience, current-voltage characteristics post-enhancement are given on both semi-logarithmic scales (b, d) and linear scales (c, f).}
\end{figure*}

\end{document}